\documentstyle[sprocl,epsf,epsfig]{article}
\def\beq{\begin{equation}}
\def\eeq{\end{equation}}
\def\bea{\begin{eqnarray}}
\def\eea{\end{eqnarray}}
\def\bq{\begin{quote}}
\def\eq{\end{quote}}

\def\APP{{\it Acta Phys.Pol.} }

\def\MPL{{\it Mod.Phys.Lett.} }

\def\NC{{\it Nuovo Cimento} }
\def\NP{{\it Nucl.Phys.} }
\def\PL{{\it Phys.Lett.} }
\def\PR{{\it Phys.Rev.} }
\def\PRL{{\it Phys.Rev.Lett.} }

\def\PTP{{\it Progr.Theor.Phys.} }

\def\ZP{{\it Z.Phys.} }

\def\gappeq{\mathrel{\rlap {\raise.5ex\hbox{$>$}} {\lower.5ex\hbox{$\sim$}}}}

\def\lappeq{\mathrel{\rlap{\raise.5ex\hbox{$<$}} {\lower.5ex\hbox{$\sim$}}}}

\begin{document}
\pagestyle{empty}
\begin{flushright} {CERN-TH/96-210}\\
{hep-ph/9611254}\\
\end{flushright}
\vspace*{5mm}
\begin{center} {\bf CURRENT ISSUES IN THE PHENOMENOLOGY}\\ {\bf  OF PARTICLE
PHYSICS}  \\
\vspace*{1cm}  {\bf John ELLIS}\\
\vspace{0.3cm} Theoretical Physics Division, CERN \\ CH - 1211 Geneva 23 \\
\vspace*{2cm}   {\bf ABSTRACT} \\ \end{center}
\noindent The present status of the Standard Model and its experimental tests are
reviewed, including indications on the likely mass of the Higgs boson. Also
discussed are the motivations for supersymmetry and grand unification, searches
for sparticles at LEP, neutrino oscillations, and the prospects for physics
at the LHC.
\vspace*{1.2cm}

\begin{center} {\it Invited plenary talk presented at the}\\ {\it Inaugural
Conference of the}\\ {\it Asia-Pacific Center for Theoretical Physics}\\ {\it
Seoul, Korea, June 1996} 
\end{center}
\vspace*{1cm}


\begin{flushleft} CERN-TH/96-210 \\ November 1996
\end{flushleft}
\vfill\eject

\setcounter{page}{1}
\pagestyle{plain}

\author{John Ellis }
\address{Theoretical Physics Division, CERN, CH - 1211 Geneva 23}
\title{Current Issues in the Phenomenology of Particle Physics\footnote{Invited
plenary talk presented at the Inaugural Conference of the Asia-Pacific Center
for Theoretical Physics, Seoul, Korea, June 1996.}}

\maketitle
\begin{abstract} The present status of the Standard Model and its experimental
tests are reviewed, including indications on the likely mass of the Higgs boson.
Also discussed are the motivations for supersymmetry and grand unification,
searches for sparticles at LEP, neutrino oscillations, and the prospects for
physics at the LHC.
\end{abstract}

\section{Introduction to the Standard Model and its Deficiencies}

\par The building blocks of the Standard Model~\cite{WS}  of particle physics
are  listed in Table 1. The fundamental electromagnetic, weak, strong and
gravitational forces are carried by the photon $\gamma$, the 
$W^{\pm}$ and $Z^0$, the gluon and (we firmly believe) the graviton,
respectively. Of these, the $\gamma$, gluon and graviton are thought to be
massless, whilst the $W^{\pm}$ and the $Z^0$ are as heavy  as medium-sized
nuclei: $80.356 \pm 0.125$ and $91.1863 \pm 0.0020$ GeV,
respectively~\cite{LEPEWWG}, leading to the very short $ \simeq 10^{-16}$ cm
range of the weak forces, as opposed to the very large and probably infinite 
ranges of the electromagnetic forces. One of the greatest issues in particle
physics - which will be discussed extensively in this talk - is to understand why
the $W^{\pm}$ and $Z^0$ intermediate bosons behave so differently from their
peers, although their basic properties, such as spins and couplings to matter
particles, seem so similar.

\begin{table}
\caption{Particles in the Standard Model}
\begin{center}
\begin{tabular}{|c|c|c|c|}     \cline{1-3}  
Gauge Boson & Mass & Range of Force &\multicolumn{1}{c}{} \\
\cline{1-3} &&&\multicolumn{1}{c}{}\\
Photon ($\gamma$) & 0 & $>10^{21}$ cm &\multicolumn{1}{c}{}  \\
$W^\pm$ & 80.356(125) GeV & $\sim 10^{-16}$ cm&\multicolumn{1}{c}{}  \\
$Z^0$ & 91.1863(20) GeV & $\sim 10^{-16}$ cm &\multicolumn{1}{c}{}  \\
gluon (g) & 0 & $\sim 10^{-13}$ cm &\multicolumn{1} {c}{} \\ 
&&&\multicolumn{1}{c}{}\\  \hline
Lepton & Mass & Quark & Mass \\ \hline
&&&\\
$\left(\matrix{e^-\cr \nu_e}\right)$ & $\matrix{1/2~{\rm MeV}\cr < 1 ~{\rm
eV}}$ & $\left(\matrix{u\cr d}\right)$ & $\matrix{\sim 5~{\rm MeV} \cr \sim 8~{\rm
MeV}}$
\\
&&&\\
$\left(\matrix{\mu^-\cr \nu_\mu}\right)$ & $\matrix{105~{\rm MeV}\cr < 0.2~{\rm
MeV}}$ & $\left(\matrix{c\cr s}\right)$ & $\matrix{1.5~{\rm GeV} \cr \sim 100~{\rm
MeV}}$
\\
&&&\\
$\left(\matrix{\tau^-\cr \nu_\tau}\right)$ & $\matrix{1.78~{\rm GeV}\cr < 23~{\rm
MeV}}$ & $\left(\matrix{t\cr b}\right)$ & $\matrix{172\pm 6~{\rm MeV} \cr 5~{\rm
GeV}~?}$
\\
&&&\\
\hline
\end{tabular}
\end{center}\end{table}

\par The fundamental particles of matter, the quarks and leptons, are also
listed in Table 1, together with the information we currently possess concerning
their masses. As we shall discuss in more detail later on,  experiments at LEP
have told us that there can be no more than three light neutrino
species~\cite{LEPEWWG},  and hence presumably no more than three charged leptons
and three corresponding pairs of quarks. Thus, the discovery last  year of the
top quark $t$~\cite{tquark} would appear to complete the Mendeleev table of
elementary particles. A basic issue is to understand the number and variety of
types of fundamental matter particle, as well as the range of their masses. As
you see from Table 1, currently we only have upper limits on the neutrino
masses~\cite{PDG}. Another of the major current issues in particle physics to be
discussed later is whether they are strictly zero, as suggested by the Standard
Model, or are non-zero as suggested by most attempts to grand unify the strong,
weak and electromagnetic interactions.

\par Experiments at LEP and elsewhere have by now tested the Standard Model, as
summarized in Table 1, down to levels from $1 \%$ to $1$ per mille. Althought
there have been a few `anomalies' to be discussed in more detail later, there
are no confirmed accelerator data that disagree with the Standard Model.
Nevertheless, particle theorists are convinced that it is incomplete, for
reasons that we now review briefly. First of all, the Standard Model contains at
least $19$ parameters: $3$ gauge couplings 
$g_{1,2,3}$ for the $U(1), SU(2)$ (electroweak) and $SU(3)$ (strong) factors in
the Standard Model gauge group, and $1$ CP-violating vacuum phase angle
$\theta_3$ for the strong interactions; $6$ quark masses, $3$ lepton masses and
$4$ parameters to describe the couplings of the $W^{\pm}$ to quarks; and the
masses of the $W^{\pm}$ and the Higgs Boson $H$ (of which more later). The
issues motivating theoretical attempts understand these many parameters by going
beyond the Standard Model can be collected into the following $3$ categories.

\par {\bf The Issue of Unification}: Can the disparate fundamental forces listed
in Table 1 be regarded as different aspects of a single Grand Unified Theory
(GUT)? This could have observable implications for proton decay as
well as neutrino masses, and predicting testable relations between the Standard
Model gauge couplings $g_i$ and between quark and lepton masses.

\par {\bf The Issue of Flavour}: Why are there so many different types of quarks
and leptons, and what explains their mixing and CP violation? Some suggest that
this might reflect a new level of compositeness within the matter particles, a
speculation revived recently in connection with data from the Fermilab ${\bar p}
p$ collider~\cite{CDFET}.

\par {\bf The Issue of Mass}: What is the origin of the particle masses? Is it
the Higgs Boson postulated in the Standard Model? If so, why are the masses of
the Standard Model particles so much smaller than the Planck Mass $M_P \simeq
10^{19}$ GeV~\cite{hierarchy},  which is the only candidate we have for a
fundamental mass scale in physics? Is this hierarchy of masses protected by
supersymmetry?

\par All of these issues should be resolved within the {\bf Theory of Everything}
(TOE), which should also include gravity and reconcile it with  quantum
mechanics, explain the origin of space and time and why there  are just $4$
large dimensions, etc.. The only candidate we have is superstring theory, which
is discussed here by Gross~\cite{gross}. My r\^ole at this meeting is to address
the previous issues, which we start by examining the bedrock of the Standard
Model.

\section{Testing the Standard Model at LEP and Elsewhere}

\par Experiments to test the Standard Model have been carried out over a large
range of energies and distance scales, from measurements of parity violation in
atoms at effective momentum transfers
$Q^2 \simeq 10^{-10}$ GeV$^2$, through experiments scattering leptons ($e, \mu,
\nu$) on fixed nucleon targets at $Q^2 \simeq 1$ to
$100$ GeV$^2$, to $e^+ e^-$, ${\bar p} p$ and $e p$ collider experiments at $Q^2
\simeq 10^4$ GeV$^2$. Of these, the most precise so far have been those carried
out in $e^+ e^-$ annihilation into $Z^0$ particles at LEP (at CERN) and the  SLC
(at SLAC). In particular, the largest accelerator in the world is LEP with a
circumference of $\simeq 27$ km (more on this later), whose energy has recently
been upgraded from
$\simeq 90$ GeV around the $Z^0$ peak, first to $130/140$ GeV at the end of
$1995$ (called LEP 1.5), then to $161$ GeV in mid-$1996$ (called LEP 2W), and to
$172$ GeV in late $1996$.

\par Annihilation through the $Z^0$ produces what is perhaps the most perfect
Breit-Wigner peak ever seen. The basic measurements made  at the $Z^0$ peak
include the following~\cite{Yellow}. The  {\bf Total Hadronic Production  Cross
Section} is given at the classical level by
\begin{equation}
\sigma_h^0 \, = \, {12 \pi \over M_Z^2}  {\Gamma_e \Gamma_h \over \Gamma_Z^2}
\label{sigtot}
\end{equation} where $\Gamma_{e,h}$ are the widths for $Z^0$ decays into $e^+
e^-$ and hadrons, respectively, and
\begin{equation}
\Gamma_Z \, = \Gamma_e + \Gamma_{\mu} + \Gamma_{\tau} + N_{\nu} 
\Gamma_{\tau} + \Gamma_h
\label{gamtot}
\end{equation} is the {\bf Total Decay Rate} of the $Z^0$. Quantum (radiative)
corrections reduce the total cross section (\ref{sigtot}) by tens of
$\%$, but are now calculated with precisions at the per mille
level~\cite{Yellow}. Other important measurements are those of the  {\bf
Leptonic Partial Decay Rates} $\Gamma_{\ell} = \Gamma_{e,\mu, \tau}$, which are
equal in the Standard Model, via the ratios
\begin{equation} R_{\ell} \, = \, {\Gamma_h \over \Gamma_{\ell}}
\label{rl}
\end{equation} A measurement that has ignited considerable interest during the
past year has been that of the {\bf Partial Decay Rate for
$Z^0$ decay into ${\bar b} b$}, parametrized by
\begin{equation} R_b \, = \, {\Gamma_b \over \Gamma_h}
\label{rb}
\end{equation} A comparison of the measurements of all the visible $Z^0$ decays 
with  that of $\Gamma_Z$ (\ref{gamtot}) via (\ref{sigtot}) enables the {\bf
Invisible $Z^0$ Decay Width}
\begin{equation}
\Gamma_{inv} \, = \, N_{\nu} \Gamma_{\nu}
\label{gaminv}
\end{equation} to be measured, and hence, since $\Gamma_{\nu}$ can be calculated
very precisely in the Standard Model, the number of equivalent light neutrino
species $N_{\nu}$.

\par In addition to these cross-section measurements, there are also precision
determinations of the {\bf Forward-Backward Asymmetries}
\begin{equation} A_{FB} \, = \, {\int_0^1 d({\rm cos}\theta) {d \sigma \over {\rm
cos}\theta} - \int_{-1}^0 d({\rm cos}\theta) {d \sigma \over {\rm cos}\theta}
\over \sigma}
\label{asym}
\end{equation} for the various flavours of leptons and quarks, where $\theta$ is
the polar angle relative to the incoming $e^-$ beam, as well as the {\bf
Final-State $\tau$ Polarization}. It is also possible to measure the {\bf
Polarized-Beam Asymmetry} $A_{LR}$, defined as the difference in cross sectioxns
for left- and right-polarized
$e^-$ beams:
\begin{equation} A_{LR} \, = \, {\sigma_L - \sigma_R \over \sigma_L + \sigma_R}
\label{alr}
\end{equation} if one has longitudinally-polarized beams, as at the
SLC~\cite{ALR}. LEP only has transversely-polarized beams, which are useful in
their own way, as we shall see shortly.

\par The current set of precision high-energy electroweak measurements at LEP,
the SLC and the Fermilab ${\bar p} p$ collider is shown in Table 2, and I shall
now comment on some of the most interesting items on the list~\cite{LEPEWWG}.
Most basic of all is the measurement of the $Z^0$ mass from LEP. This requires
very precise calibration of the LEP beam energy, which is provided by resonant
destruction of the transverse beam polarization. In order to obtain the stated
precision,  which is comparable to the accuracy with which the Fermi weak
coupling is measured in $\mu$ decay, a myriad of delicate effects such as the
temperature and humidity of the LEP tunnel must be taken into account. Effects
have also been seen which are due to the tides, which expand and contract the
rock in which LEP is embedded, altering the circumference of the
machine~\cite{tides}.  Because of the RF tuning of LEP, these alterations cause
the  orbits of the beams to move relative to the LEP magnets, which in turn
alters the beam energies by several MeV, as seen in Fig.~1(a). There are other
effects that can alter the circumference of LEP, and  hence the beam energy. One
is whether it has been raining: if the water table in the Jura mountains rises,
the absorbent rock expands, as seen in Fig.~1(b). Another is the water level in
Lake Geneva:  each Spring, lake water is let out to make room for molten snow
from the Alps. When the weight of the lake water is removed, the rock rises and
expands with a time delay of about $100$ days, as seen in
Fig.~1(c)~\cite{water}. All of these variations have been taken into account in
the calculation of $M_Z$ shown in Table~2.

\begin{table}
\caption{High-Energy Precision Electroweak Data Set}
\begin{center}
\begin{tabular}{|ccc|cc|}     \hline
&&&&\\ 
$M_Z$ & 91.1863  (20) & GeV & $A^b_{FB}$ & 0.0979  (23) \\ 
&&&&\\
$\Gamma_Z$ & 2.4946  (27) & GeV & $A^c_{FB}$ & 0.0733  (49) \\
&&&&\\
$\sigma^0_n$ & 41.508  (56) & nb & $\sin^2\theta_{eff}(Q_{FB})$ & 0.2320  (10)
\\&&&&\\
$R_L$ & 20.778  (29) & & $M_W$ & 80.356  (125) \\
&&&&\\
$A^L_{FB}$ & 0.0174  (10) & & $\sin^2\theta_W(A_{LR})$ & 0.23061  (47) \\
&&&&\\
$A_\tau$ & 0.1401  (67) & & $R_b$ (SLC) & 0.2149  (38) \\
&&&&\\	$A_e$ & 0.1382  (76) & & $A_b$ (SLC) & 0.863  (49) \\
&&&&\\
$R_b$ & 0.2179  (12) & & $A_c$ (SLC) & 0.625  (84) \\
&&&&\\
$R_c$ &  0.1715  (56) &  &&\\
&&&&
 \\ \hline
\end{tabular}
\end{center}\end{table}

\par Another effect first surfaced as variations in the LEP energy during fills,
which diminished during the night~\cite{TGV}. These were eventually traced to
nearby electric trains: see the passage of a TGV in Fig.~1(d). The explanation is
that some of the return current passes through the Earth, and in particular
through LEP (which is a relatively good conductor). Since the LEP beam energy
shifted during the course of each fill, and since the beam was calibrated at the
ends of the fills, there was a systematic correction to the beam energy and hence
$M_Z$ of a few MeV, which has now been taken into account in the value quoted in
Table~2~\cite{LEPEWWG}.

\begin{figure}
\hglue.3cm
\mbox{\epsfig{figure=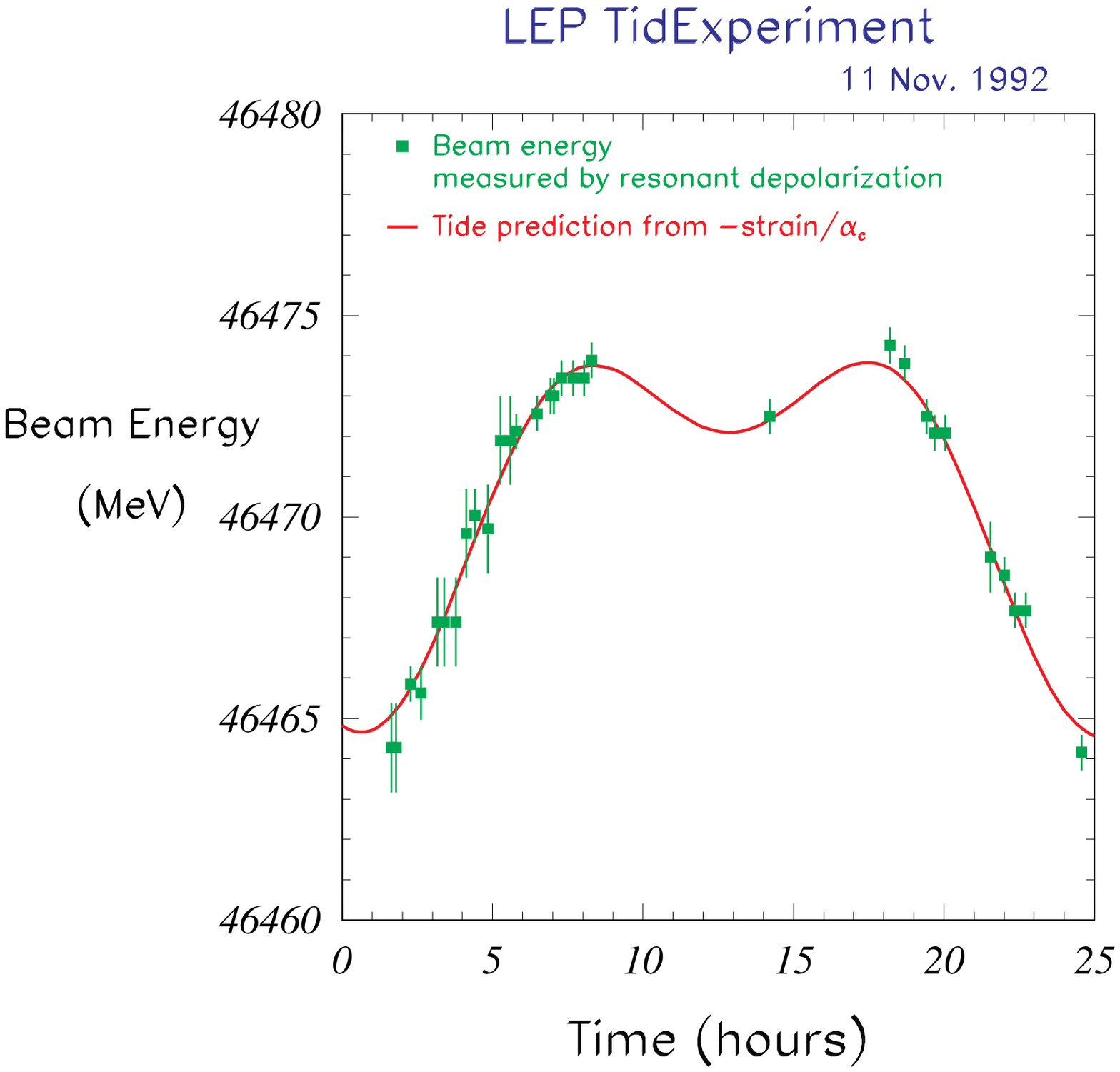,width=5cm}(a)
\epsfig{figure=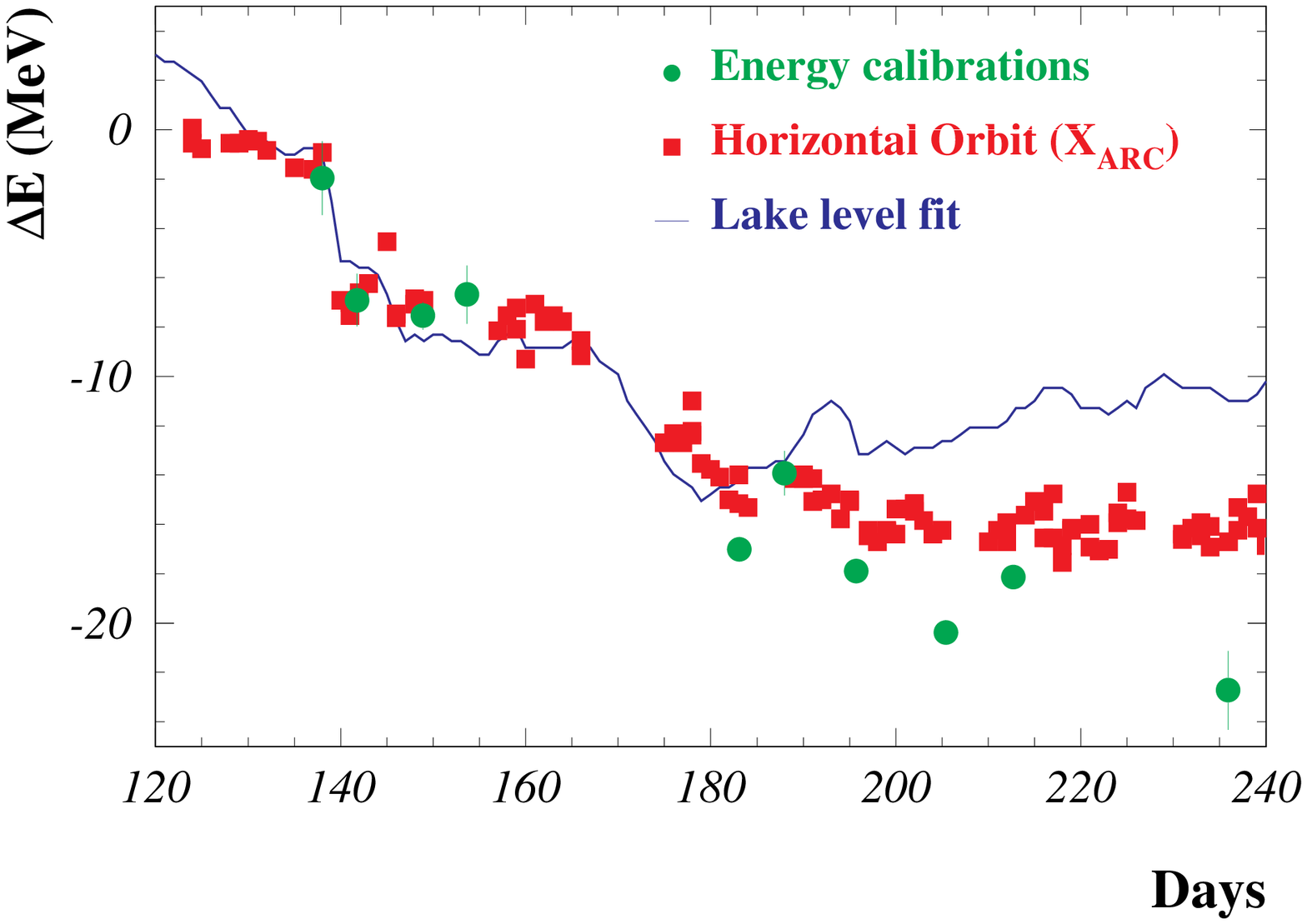,width=5cm}(c)}
\hglue.3cm
\mbox{\epsfig{figure=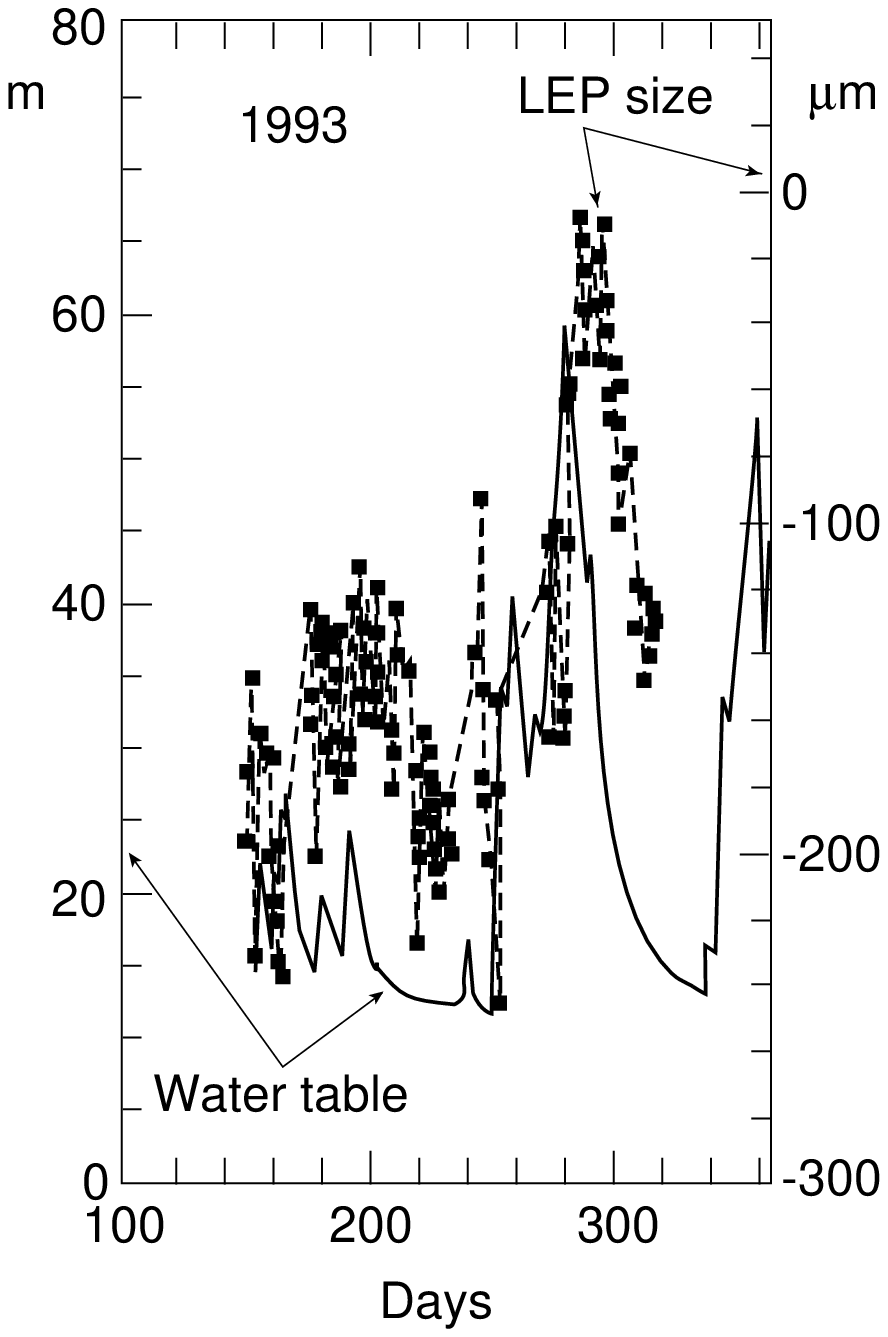,width=5cm}(b)
\epsfig{figure=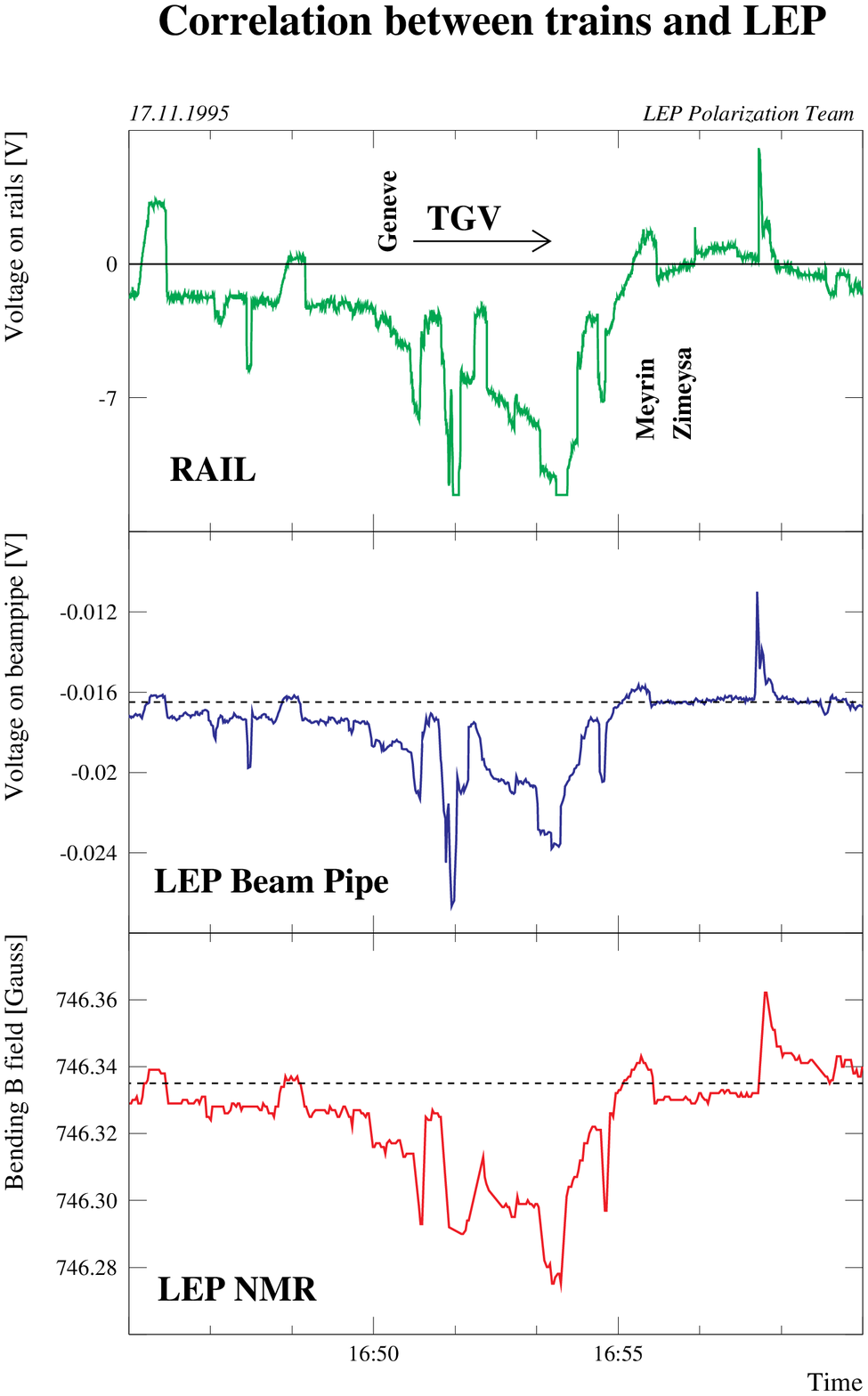,width=5cm}(d)}
\caption[]{Sensitivity of the LEP beam energy to (a) tides \cite{tides}: the
solid  lines are due to a tidal model, (b) the water table in the Jura mountains 
and (c) the level of Lake Geneva \cite{water}, and  (d) the ``TGV effect" on the
LEP beam energy \cite{TGV}.
}
\end{figure}

\par Turning now to some of the other measurements in Table~2, the relative beam
energy calibrations are also important for the determination of $\Gamma_Z$, but
the TGV effect is not so important, since it tends to move all the beam energies
by similar amounts. Last year, measurements of $R_b$ and $R_c$ appeared to come
into significant disagreement with the Standard Model predictions, but this
anomaly now seem to have evaporated, as we discuss in section 5. In the previous
year, there had been some concern about the compatibility of the LEP and SLC
measurements of the electroweak mixing angle sin$^2 \theta_W$, but this no longer
seems to be a significant discrepancy~\cite{LEPEWWG},\cite{ALR}. Finally, note the
accuracy with which the effective number of light $\nu$ species has now been
measured~\cite{LEPEWWG}:
\begin{equation} N_{\nu} \, = \, 2.989 \pm 0.012
\label{nnu}
\end{equation} I had always hoped $N_{\nu}$ would turn out to be non-integer,
say $\pi$ or, even better, $e$, but this was not to be. Even so, the measurement
(\ref{nnu}) is a useful constraint on supersymmetric extensions of the Standard
Model, as we discuss later.

\par What use are the precise numbers in Table~2? One answer is provided by
their sensitivity to the masses of unseen particles, via quantum (radiative)
corrections. For example, at one loop:
\begin{equation} M_W^2 \hbox{sin}^2 \theta_W = M_Z^2 \hbox{cos}^2 \theta_W
\hbox{sin}^2 \theta_W = {\pi \alpha \over \sqrt{2} G_{\mu}} (1 + \Delta r)
\label{deltar}
\end{equation} and the correction $\Delta r$ is sensitive to the masses of the
top quark and the Higgs boson. In the case of the top, the sensitivity is
quadratic~\cite{Veltman+}:
\begin{equation}
\Delta r \, \simeq \, {3 G_{\mu} \over 8 \pi^2 \sqrt{2}} m_t^2
\label{mtdep}
\end{equation} for $m_t >> m_b$. This sensitivity enables precision data to be
used to a theoretical prediction for
$m_t$. In the case of the Higgs boson,  the sensitivity is unfortunately only
logarithmic at the one-loop level~\cite{Veltman}:
\begin{equation}
\Delta r \, \simeq \, {\sqrt{2} G_{\mu} \over 16 \pi^2} M_W^2 [{11 \over 3}
\hbox{ln} ({M_H^2 \over M_W^2}) + \dots]
\label{mhdep}
\end{equation} making the use of precision data to predict $M_H$ much more
delicate. 

\par If just one quantum correction is determined, e.g., $\Delta r$ from
measurements of $M_{W,Z}$~(\ref{deltar}), a trade-off between the values of
$m_t$ and $M_H$ is possible, but these can both be determined if enough quantum
corrections are pinned down~\cite{EFL}. Our latest prediction of $m_t$, based on
a $\chi^2$  analysis of the available precision electroweak data shown
is~\cite{EFL}
\begin{equation} m_t \, = \, 157^{+16}_{-12} \, \hbox{GeV}
\label{mtpred}
\end{equation} including the error associated with leaving $M_H$ a free
parameter. This prediction of $m_t$ is compatible with the latest Fermilab
measurement~\cite{CDFD0}:
\begin{equation} m_t \, = \, 175 \pm 6 \, \hbox{GeV}
\label{mtmeas}
\end{equation} The consistency between (\ref{mtpred}) and (\ref{mtmeas})  is a
non-trivial check of the Standard Model at the quantum level. The agreement
between (\ref{mtpred}) and (\ref{mtmeas}) also means that they can legitimately
be combined to yield~\cite{EFL}
\begin{equation} m_t \, = \, 172 \pm 6 \, \hbox{GeV}
\label{mtcomb}
\end{equation} which is the current best estimate of $m_t$ within the Standard
Model.

\section{The Origin of Mass}

Let us now address in more detail the central issue of the origin of mass: How
come the $W^{\pm}$ and $Z^0$ are massive, whereas the $\gamma$ and gluon are
massless? The core of this problem lies in the fact that a massless spin-$1$
particle has only two polarization states with helicities $\pm 1$, whereas a
massive spin-$1$ particle has three polarization states: $1, 0 -1$. The
suggestion is that the primordially massless $W^{\pm}$ and $Z^0$
combine with extra spin-$0$ particles which provides their third polarization
states, enabling them to become massive. As we shall discuss in more detail
shortly, realistic electroweak models require there to exist at least one
physical scalar, in addition to the spin-$0$ degrees of freedom `eaten' by the
massive $W^{\pm}$ and $Z^0$. there is no direct experimental evidence for such a 
Higgs boson, and searches at LEP have established the lower limit
\begin{equation} M_H \, \ge \, 66 \, \hbox{GeV}
\label{lowermh}
\end{equation} There are, however, theoretical arguments based on unitarity,
which suggest that~\cite{upper}
\begin{equation} M_H \, \le \, 1 \, \hbox{TeV}
\label{uppermh}
\end{equation} and hence that the Higgs boson may lie within the reach of the
next generation of accelerators, such as the LHC discussed in section 7.

\par At a more theoretical level, the problem of mass can be seen as requiring a
`breakdown' of the electroweak gauge symmetry. To avoid non-renormalizability
problems in higher-order calculations, this `breakdown' should only be
spontaneous, i.e., it should be due to the condensation in the electroweak
vacuum of some field
$X$ with non-zero electroweak isospin:
\begin{equation} M_{W,Z} \ne 0 \,\, <=> \, \, <0|X_{I, {I_3}}|0> \ne 0
\label{spont}
\end{equation} The numerical values of the $W^{\pm}$ and $Z^0$ masses indicate a
particular ratio
\begin{equation}
\rho \, = \, {M_W^2 \over M_Z^2 \hbox{cos}^2 \theta_W} \simeq 1
\label{rhoratio}
\end{equation} which corresponds to the simplest choice $I = 1/2$~\cite{RV}.
This is  what is required also to give masses to the quarks and leptons, and was
the choice of Weinberg and Salam when they originally wrote down  the Standard
Model~\cite{WS}.

\par The next Big Question is whether this field $X$ sitting in the vacuum is
elementary (as Weinberg and Salam postulated) or composite. The latter
possibility may be appealing to many of you from a condensed-matter background,
who are familiar with the role of Cooper pairs in superconductivity and pairing
in superfluid
${}^3$He, and could also be reminiscent of quark condensation in QCD: $<0|{\bar
q} q|0> \ne 0$. Composite Higgs models have included ${\bar t} t$ condensate
models~\cite{Nambu} - but these initially wanted $m_t > 200$ GeV, and are now
looking for epicycles - and so-called technicolour models~\cite{TC}, which
postulate new fermions bound by new interactions that become strong on an energy
scale around $1$ TeV. At least the simplest versions of such models seem to be
disfavoured by a 
$\chi^2$ analysis of the precision electroweak measurements~\cite{EFLTC}, as
seen in Fig.~2, and variations in these models are now also being explored.

\begin{figure}
\hglue1cm
\epsfig{figure=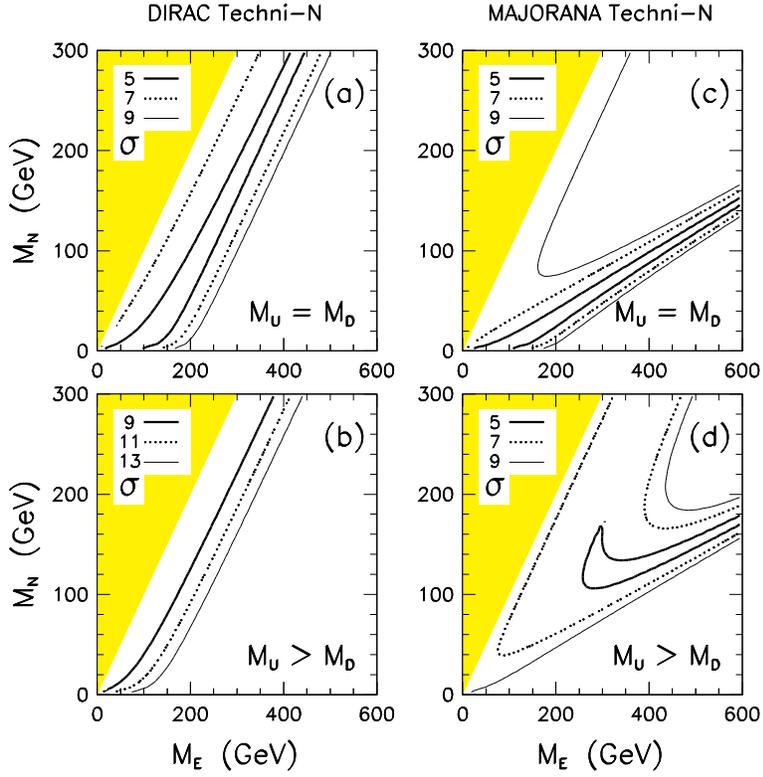,width=10cm}
\caption[]{
Contours \cite{EFLTC} of $\sigma\equiv\sqrt{\Delta\chi^2}$ for one-generation 
models with either
Dirac technineutrinos (a), (b) or a Majorana technineutrino (c), (d). Note that 
$\sigma\gappeq$ 4.3 in all of
the TC parameter space, to be compared with $\sigma$ = 2.6 in the SM at the 
reference point $(m_t$ = 170 GeV,
$M_H = M_Z$). In the case of techniquark mass degeneracy $(M_U = M_D)$, however,
the Dirac model becomes highly disfavoured. In all cases, $\xi$ = 1/2 is assumed.}
\end{figure}

\begin{figure}
\hglue2.5cm
\epsfig{figure=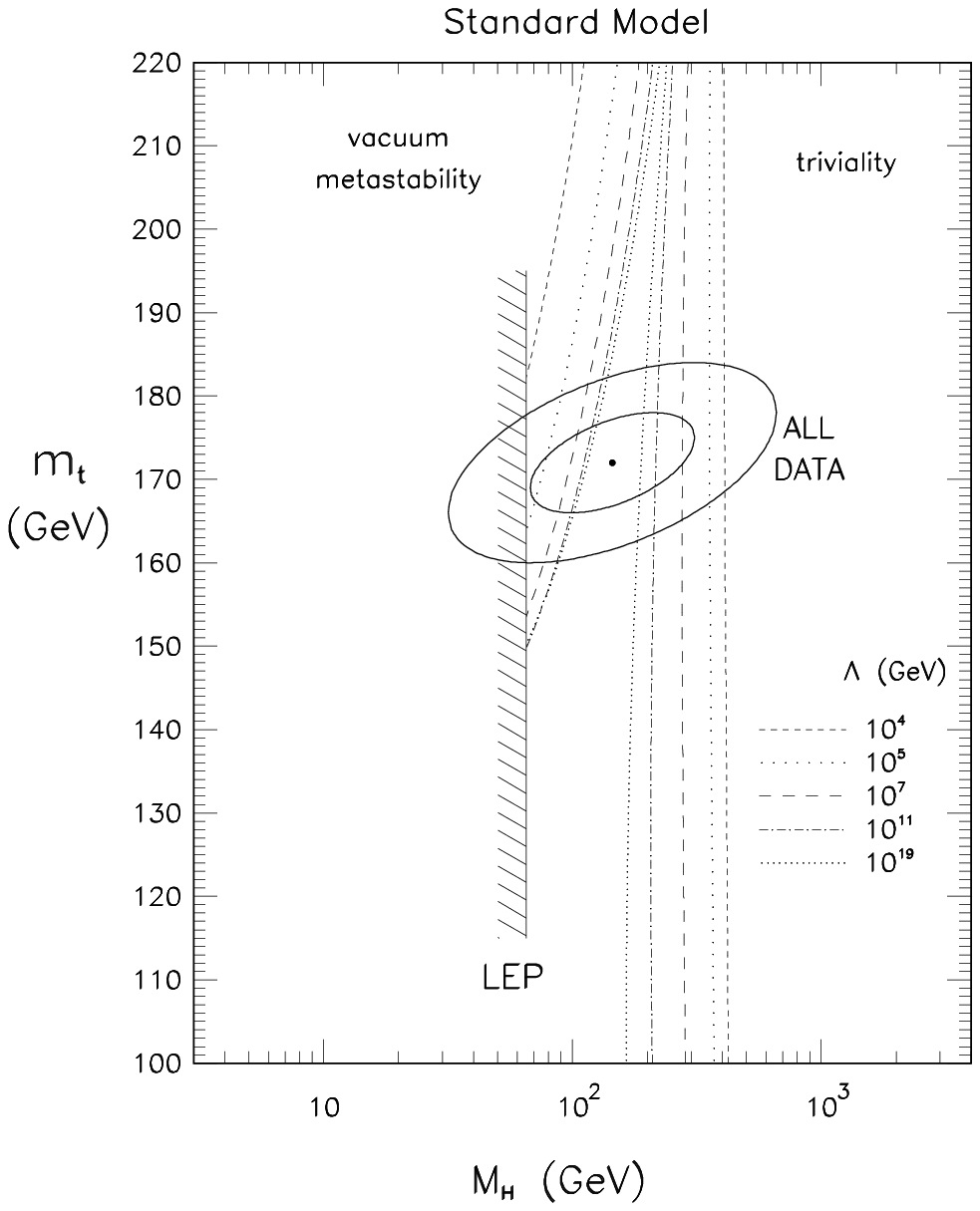,width=7cm}
\caption[]{
Indirect bounds \cite{EFL} on $(M_H,\,m_t)$ and one-sided experimental
		and theoretical limits in the Standard Model.  The solid 
		ellipses represent the 1-$\sigma$ and 2-$\sigma$ contours 
		from the best-fit  Gaussian distribution obtained by 
		analysing all electroweak precision data, including the 
	measurement of $m_t$ at CDF and D0. The hatched line is
		the LEP lower bound on $M_H$ \protect. 
		The other curves represent the lower and upper limits on 
		$M_H$ from vacuum metastability  \protect and 
		triviality \protect respectively, as 
		functions of the scale of new physics $\Lambda$.
}
\end{figure}

\par In the absence so far of a credible composite alternative, we are led to
examine more closely the elementary possibility. If there is just a single
$I=1/2$ Higgs doublet, consisting of two complex fields, the three degrees of
freedom eaten by the $W^{\pm}$ and $Z^0$ leave behind a single physical higgs
boson to be detected. A $\chi^2$ analysis of precision electroweak data within
the minimal Standard Model with just this one physical elementary Higgs boson
leads to the estimate~\cite{EFL}, see also~\cite{LEPEWWG}:
\begin{equation} M_H \, = \, 145 {\,}^{+164}_{-77} \, \hbox{GeV}
\label{mhest}
\end{equation} 
Figure 3 displays $\Delta \chi^2 = 1,4$ contours in the
$M_H, m_t$ plane, for a fit~\cite{EFL} to the precision electroweak data which
includes the direct CDF and D0 measurements of $m_t$. These results are quite
consistent with the idea of a weakly-coupled elementary Higgs boson within reach
of planned accelerators.

\section{Motivations for Supersymmetry}

\par There are, however, theoretical problems with such a simple possibility,
associated with the gross disparities in the known mass scales in physics. The
only candidate we have for a fundamental mass scale is the Planck mass, related
to Newton's constant: $M_P = 1/\sqrt{G_N}$. Why is $M_W \ll M_P$? This is
commonly known as the Hierarchy Problem~\cite{hierarchy}, which can be rephrased
as: why is $G_F \gg G_N$? Some atomic and condensed-matter physicists may
consider this question  remote from their concerns, but it is equivalent to the
question: why is the Coulomb potential in an atom so much smaller than the
Newtonian potential, i.e., why is $e^2 \gg G_N m^2$, where $m$ is a typical
particle mass?

\par These questions are particularly worrying for models with an elementary
Higgs boson, because its mass is subject to large quantum corrections, meaning
that the small physical value (\ref{uppermh}) can be obtained only at the
expense of extreme fine tuning. One usually prefers that quantum corrections to
a measurable quantity not be much larger than its physical value, since
otherwise its value would seem unnatural~\cite{hierarchy}. An example of a
physical quantity with naturally small quantum corrections is a fermion mass:
\begin{equation}
\delta m_f \, \simeq \, ({\alpha \over \pi}) m_f \, \hbox{ln}  ({\Lambda \over
m_f}) 
\label{deltamf}
\end{equation} which is not much greater than $m_f$ for any plausible value of
the cutoff $\Lambda \le M_P$.

\par The same cannot be said for quantum corrections to the mass of an
elementary Higgs boson, which are quadratically divergent:
\begin{equation}
\delta M_H^2 \simeq g^2_{f,W,H} \int^{\Lambda} {d^4 k \over (2 \pi)^4} {1 \over
k^4} \simeq ({\alpha \over \pi})
\Lambda^2 >> M_H^2
\label{deltamh}
\end{equation} If one inserts a guess for the cutoff $\Lambda \simeq M_P$ or 
$M_{GUT}$ up to which the Standard Model may be valid, one gets a correction
which is many orders of magnitude greater than the possible physical value of
$M_H$.

\par This unpleasant conclusion can be avoided by observing that  the fermion
and boson loops have opposite signs:
\begin{equation}
\delta M_{W,H}^2 \simeq - ({g_F^2 \over 4 \pi^2}) (\Lambda^2 + M_F^2) + ({g_B^2
\over 4 \pi^2}) (\Lambda^2 + M_B^2)
\label{cancel}
\end{equation} The leading quadratic divergences will therefore cancel if there
are equal numbers of bosons and fermions: $N_B = N_F$, and if their couplings
are identical: $g_B = g_F$. These are the conditions that a field theory
manifest supersymmetry~\cite{susy}.  After the cancellation which it enforces,
the remainder 
\begin{equation}
\delta M_{W,H}^2 \simeq ({\alpha \over \pi})|M_B^2 - M_F^2|
\label{remainder}
\end{equation} will be small, rendering the hierarchy natural, if
\begin{equation} |M_B^2 - M_F^2| \, \le \, 1 \, \hbox{TeV}^2
\label{susynear}
\end{equation} It is this squared-mass difference that should be interpreted as
the cut off $\Lambda^2$ at which new physics modifies the Standard Model.
Although there are other arguments for  supersymmetry\footnote{In particular,
safeguarding the hierarchy in some GUTs benefits from the absence~\cite{nologs}
of certain logarithmic divergences in supersymmetric theories.}, such as its
intrinsic beauty and its necessity for the consistency of string theory, this is
the only argument to indicate that supersymmetry should appear at an accessible
mass scale. It should be emphasized that this argument is, nevertheless,
qualitative and a matter of taste: even an unnatural theory may be
renormalizable. Mathematically, all one needs for renormalizability is that all
$\Lambda$ cutoff dependence can be absorbed by a finite set of bare parameters:
{\it a priori} there is no need for the bare parameters and the quantum
corrections to be comparable, as is implied by the naturalness argument. The
latter is a physical argument motivated by the absence of fine tuning, not a
precise mathematical requirement.

\par The first question you might ask is whether any of the known fermions
(quarks, leptons) could be the supersymmetric partners of any of the known
bosons ($\gamma, W^{\pm}, Z^0$, Higgs, gluon). The answer is no~\cite{Fayet},
because the fermions and bosons have different internal quantum numbers, and
hence different couplings. For example, quarks are in triplets {\bf 3} of
colour, whereas the the known bosons are singlets or octets {\bf 8} of colour,
and leptons  are the only particles that carry lepton number L. One is therefore
led to introduce supersymmetric partners for all the  known particles, as shown
in Table 3. Supersymmetry is not economical in particles, though it is
economical in principle!

\begin{table}
\caption{Particles and Sparticles}
\begin{center}
\begin{tabular}{|lcclcc|}     \hline
&&&&&\\  
Name & & Spin & Sname & & Spin \\
&&&&& \\ \hline
&&&&&\\
quark & $q$ & 1/2 & squark & $\tilde q$ & 0 \\
lepton & $\ell$ & 1/2 & slepton & $\tilde\ell$ & 0 \\
photon & $\gamma$ & 1 & photino & $\tilde\gamma$ & 1/2 \\
&$Z^0$ & 1 & zino & $\tilde Z$ & 1/2 \\
& $W^\pm$ & 1 & wino & $\tilde W^\pm$ & 1/2 \\
gluon & g & 1 & gluino & $\tilde g$ & 1/2 \\
Higgs & $H^{0,\pm}$ & 0 & higgsino & $\tilde H^{0,\pm}$ & 1/2 \\ 
graviton & $G$ & 2 & gravitino & $\tilde G$ & 3/2 \\
&&&&& \\ \hline
&&&&& \\ 
\multicolumn{6}{|c|}{
The $\tilde W^\pm$ and $\tilde H^\pm$ mix to yield two mass eigenstates }\\ 
\multicolumn{6}{|c|}{called
charginos, and the $\tilde\gamma$, $\tilde Z$ and $\tilde H^0$ mix to}\\
\multicolumn{6}{|c|}{ yield four
mass eigenstates called neutralinos.} \\ &&&&& \\ \hline
\end{tabular}

\end{center}\end{table}

\par Sparticle searches at accelerators have so far been unsuccessful: the
latest limits on squark and gluino production in
${\bar p} p$ collisions~\cite{gluinomass} indicate that
\begin{equation} m_{{\tilde q}, {\tilde g}} \, > \, 200~ \hbox{GeV}
\label{qgmass}
\end{equation} and LEP limits  on slepton production~\cite{sleptonmass} indicate
that
\begin{equation} m_{\tilde \ell} \, > 70 \, \hbox{GeV}
\end{equation} Although disappointing, these limits do not yet bite far into the
expected mass range  (\ref{susynear}). Completing these searches will be the
task of future accelerators, such as the LHC discussed in section 7.

\par Of particular interest is the lightest supersymmetric particle (LSP),
denoted by $\chi$, which is expected to be stable in many models,  and is
therefore a good candidate for the Cold Dark Matter (CDM)  believed to
constitute most of the matter in the Universe. Fig.~4(a) shows the experimental
lower limit on its mass,
\begin{equation} m_{\chi} \, \ge \, 12.8 \, \hbox{GeV}
\label{aleph}
\end{equation} 
obtained by the ALEPH collaboration~\cite{alephmchi} by combining
searches at LEP $1$ and $1.5$, and assuming large slepton masses.  As discussed
in~\cite{efos}, this and other loopholes in the ALEPH analysis that may be
filled by additional theoretical and cosmological inputs, and, as also seen in
Fig.~4(a), the bound (\ref{aleph}) may be strengthened~\footnote{Preliminary LEP
2W data also enable the limit (\ref{aleph}) be strengthened to about $20$ GeV
without additional assumptions~\cite{efos'}.} to
\begin{equation} m_{\chi} \, \ge \, 21.4 \, \hbox{GeV}
\label{efos}
\end{equation} 
As can be seen in Fig.~4(b), in many models the LSP $\chi$ has a
relic cosmological density in the range of interest for cosmological
CDM~\cite{EHNOS}, and there are reasonable prospects that it could be detected
either directly or indirectly~\cite{detection}.

\begin{figure}
\hglue3.5cm
\epsfig{figure=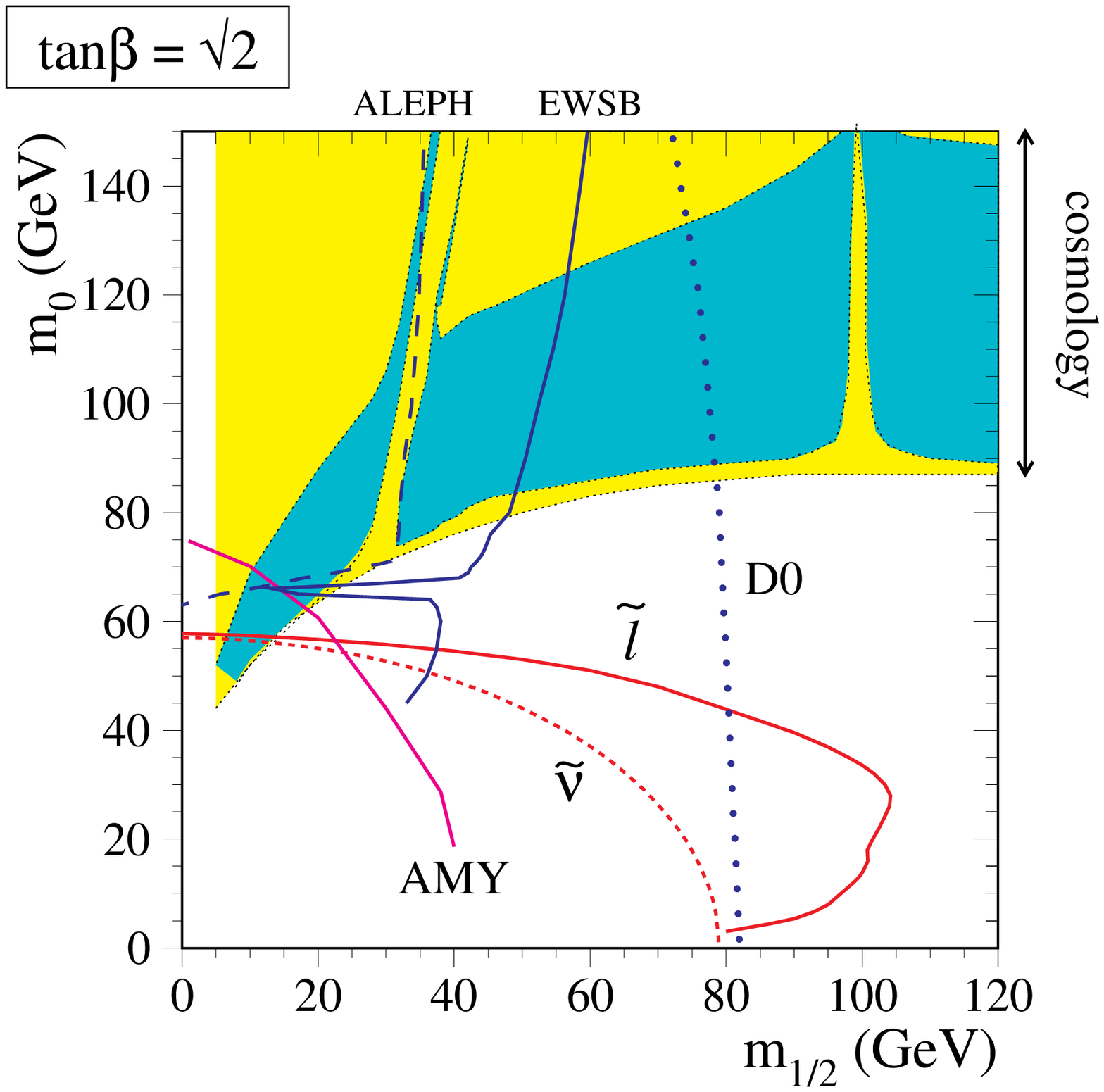,width=6cm}\\
\hglue3cm
\epsfig{figure=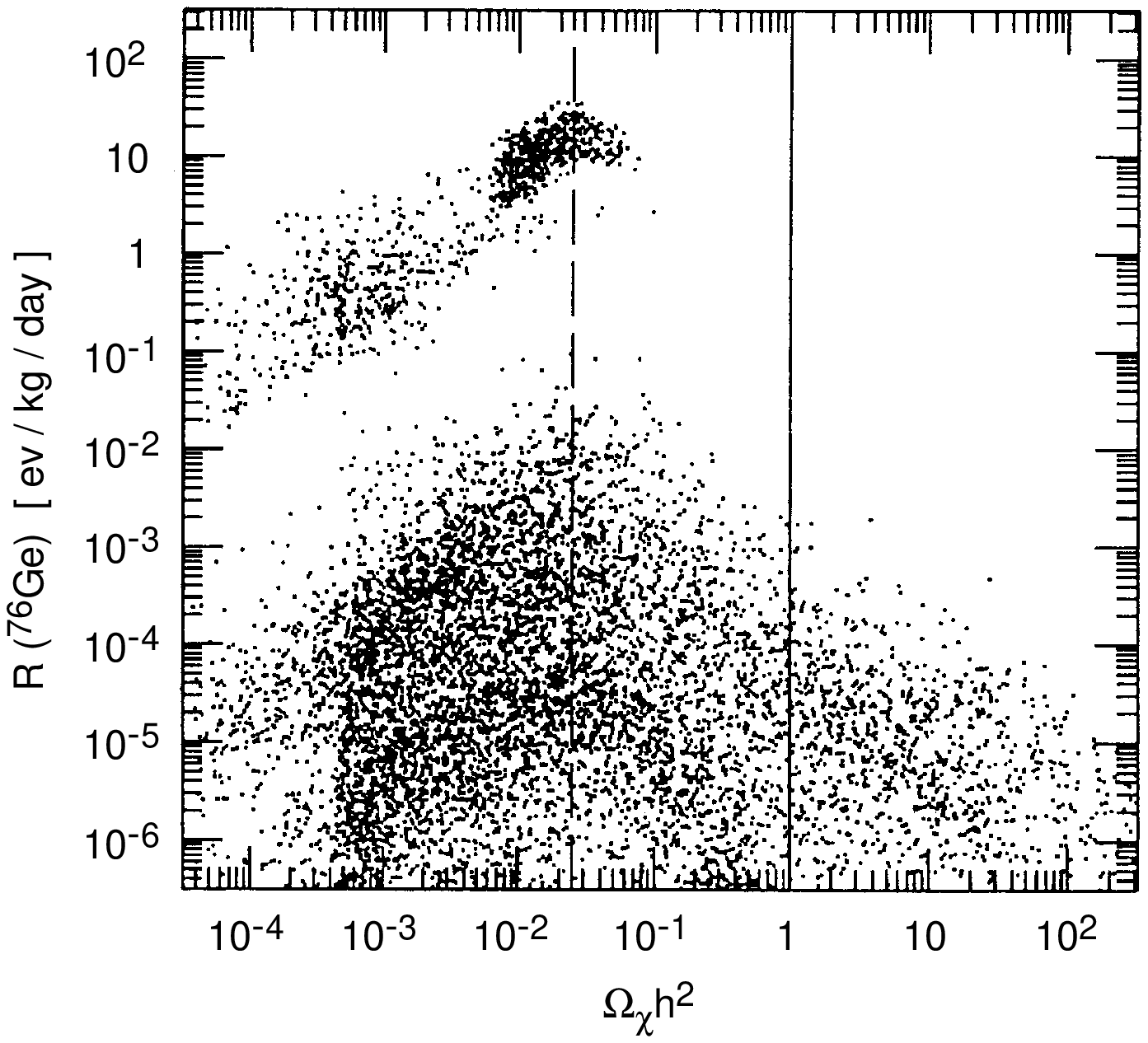,width=7cm}
\caption[]{
(a): The domain of the $(m_{1/2}, m_0)$ plane for $\mu < 0$ and
tan$\beta = \sqrt{2}$ that is excluded by ALEPH chargino and
neutralino searches~\cite{alephmchi} (long-dashed line), by the $Z^0$
limit on $m_{\tilde \nu}$ (short-dashed line), by the LEP
limits on slepton production (solid line), by
single-photon measurements (grey line), and by the D0 limit
on the gluino mass (dotted line)~\cite{efos}. The region of the plane in
which $0.1 < \Omega_{\chi} h^2 < 0.3$ for some experimentally-allowed
value of $\mu < 0$ is light-shaded, whilst the dark-shaded region is for
$\mu$ determined by dynamical EWSB. The constraint derived from the
ALEPH searches~\cite{alephmchi} when dynamical EWSB is imposed is also 
shown as a solid line~\cite{efos}.\phantom{xxxxxxxxxxxxxxxxxxxxxxxxxxxxxxxxx} \\
(b): Relic density of supersymmetric particles, calculated in a sampling
of different models~\cite{detection}, together with the estimated scattering rate
on
$^{76}$Ge.
}
\end{figure}

\par In view of all the experimental disappointments to date, are there any
experimental reasons for believing in supersymmetry? In my view, there are two
encouraging tentative indications. One is the apparent lightness (\ref{mhest})
of the Higgs boson favoured by the precision electroweak data~\cite{EFL} (see
also~\cite{LEPEWWG}). The mass of the lightest Higgs boson in the minimal
supersymmetric extension of the Standard Model can be
calculated~\cite{susyhiggs}, and it comes out below $130$ GeV or so, in high
consistency with (\ref{mhest}). The second tentative indication is furnished by
the consistency of LEP and other measurements of the Standard Model gauge
coupling strengths $\alpha_{1,2,3}$ with the prediction of a minimal
supersymmetric GUT~\cite{susygut}. As seen in the left-hand part of Fig.~5, GUTs
both with and without supersymmetry are in good qualitative agreement with the
measurements. However, when we blow up the vertical scale, as shown in the other
two parts of Fig.~5, we see that the non-supersymmetric GUT prediction disagrees
with the data, whereas the minimal supersymmetric GUT is very close.  Plausible
variations in the supersymmetric GUT model-building can bring the prediction
into perfect agreement with the data~\cite{beijing}.

\begin{figure}
\hglue2.5cm
\epsfig{figure=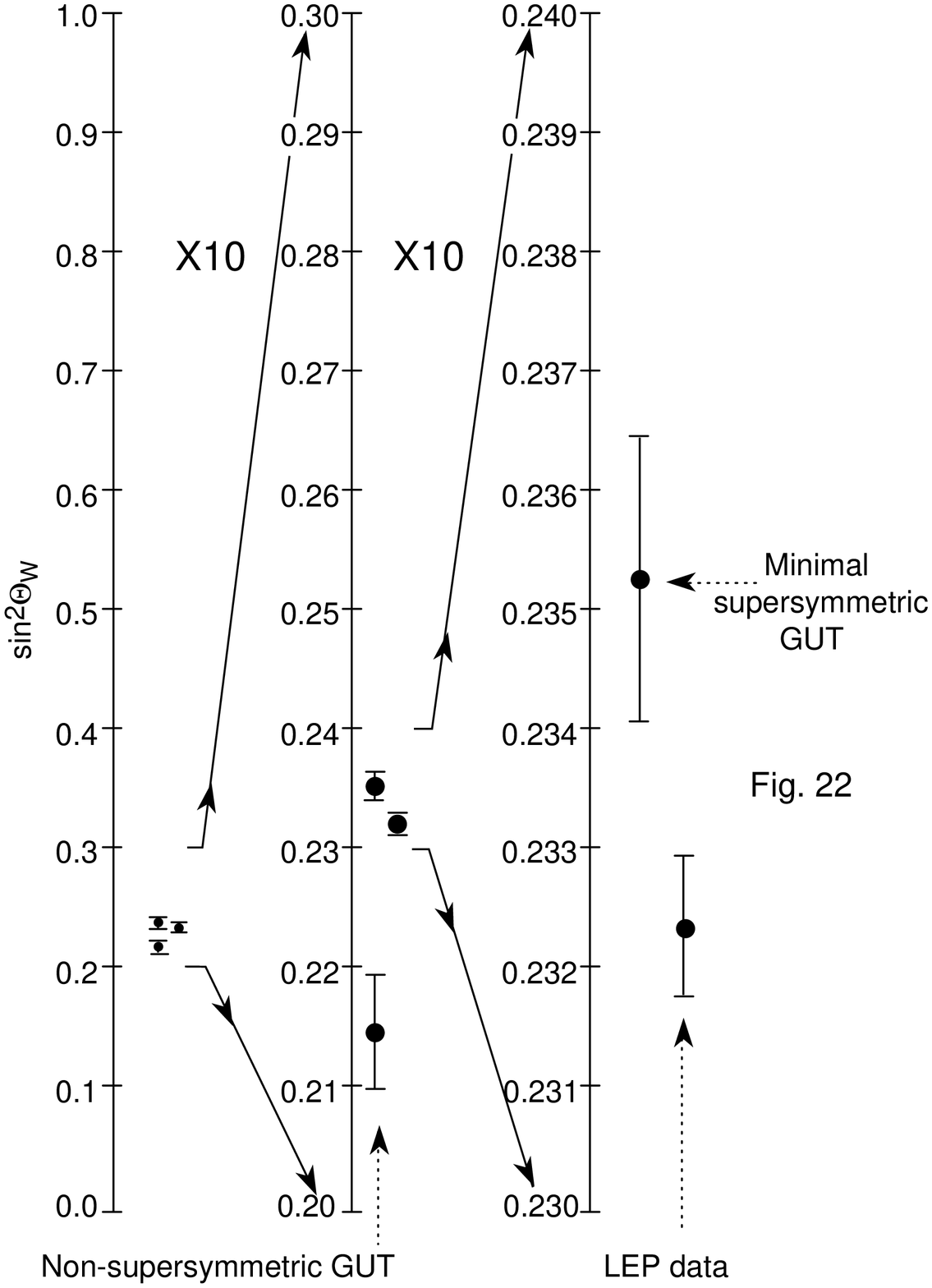,width=10cm}
\caption[]{
Gee-whizz plot showing how well GUT predictions of
$\sin^2\theta_W$ agree with the experimental data. 
}
\end{figure}

\section{Experimental Anomalies?}

Several possible experimental anomalies have been under active  discussion
during the past few months, and here are updates and  opinions on two of them.

{\bf $R_b/R_c$ `Crisis' at LEP}: This is the disagreement between the Standard
Model prediction for these two $Z^0$ decay branching ratios (\ref{rb}) mentioned
previously, which was simmering at the $2-\sigma$ level until a new round of
preliminary measurements announced in 1995 escalated the apparent discrepancies
to $3.7 \sigma$ for $R_b$ and $2.5 \sigma$ for $R_c$~\cite{LEPEWWG1995}. It
should be remembered that these are among the most difficult and complex LEP
measurements, with large systematic errors associated with the modelling of
complicated hadronic final states. One suggestion was that some misunderstanding
of these problems might responsible for the apparent anomalies. Alternatively, 
it was suggested that the discrepancies might be due to some new physics beyond
the Standard Model, such as supersymmetry~\cite{susyRb} or a new
$Z'$ boson~\cite{zprime}.

\par We studied~\cite{ELN} the former possibility, constraining possible
supersymmetric models using all the available phenomenological limits on their
parameters, including those from unsuccessful sparticle searches at LEP and
Fermilab,  the experimental lower limit on the mass of the lightest 
supersymmetric Higgs boson, and the measured rate of 
$b \rightarrow s \gamma$ decay. Particularly important was the lower limit on
the chargino mass from LEP $1.5$. Out of over $450,000$ parameter choices with 
tan$\beta < 5$ and sparticle mass parameters below $250$ GeV, we found none that
yielded a contribution to
$R_b$ greater than 0.0017, which was too small to make a significant
contribution to the resolution of the $R_b$ discrepancy. We concluded that ``...
it may be necessary to review  carefully the calculation and simulation of the
Standard Model contributions to $R_b$ ..."~\cite{ELN}.

\par Considerable new effort has now been put into the simulation of
$Z^0 \rightarrow {\bar b} b$ and ${\bar c} c$ decays, and several significant
new measurements of
$R_b$ have been announced in the Summer of 1996,  notably by the ALEPH, DELPHI
and SLD collaborations~\cite{newRb}. None of the new measurements disagrees
significantly with the Standard Model. There is still some question how to
combine these new results with the previous ones, but it seems that the
$R_b/R_c$ anomaly is on the way to resolution within the Standard
Model~\cite{LEPEWWG}.

\par {\bf High-$E_T$ jet spectrum}: this excited considerable attention in
January 1996, when the CDF collaboration revealed their spectrum of high-$E_T$
jets at the Fermilab ${\bar p} p$ collider, which exhibited for $E_T > 200$ GeV
an apparent excess above  Standard Model  calculations made with the
distributions of partons inside protons proposed previously~\cite{CDFET}.  A
possible explanation within the Standard Model was that of modifying these model
parton distributions~\cite{Durham},\cite{CTEQ}, which cannot be calculated accurately
from first principles. Possible explanations beyond the Standard Model included
the suggestion that the parton distributions should not be altered, but that the
strong coupling $\alpha_s$  might decrease more slowly than expected in standard
QCD, because of the appearance of new strongly-interacting degrees of freedom
such as squarks and gluinos~\cite{slowas}. Another suggestion was to keep the
usual parton distributions and leave QCD unscathed, but postulate an extra
interaction between quarks, due to the exchange of a new electroweak boson $Z'$,
which might also resolve the $R_b/R_c$ `crisis' mentioned above~\cite{zprime}.
Finally, the most radical interpretation was that mentioned by the CDF
collaboration~\cite{CDFET},  namely that quarks might be composite objects.

\par We analyzed in detail~\cite{ER} the possible effects of a sparticle
threshold on the jet spectrum, via their quantum corrections to parton-parton
scattering cross sections. The maximum effect that we found was just a few $\%$
in a narrow region around the sparticle threshold, as seen in Fig.~6, and we
found that previous calculations~\cite{slowas} far above and below threshold were
unreliable guides.

\begin{figure}
\mbox{\epsfig{figure=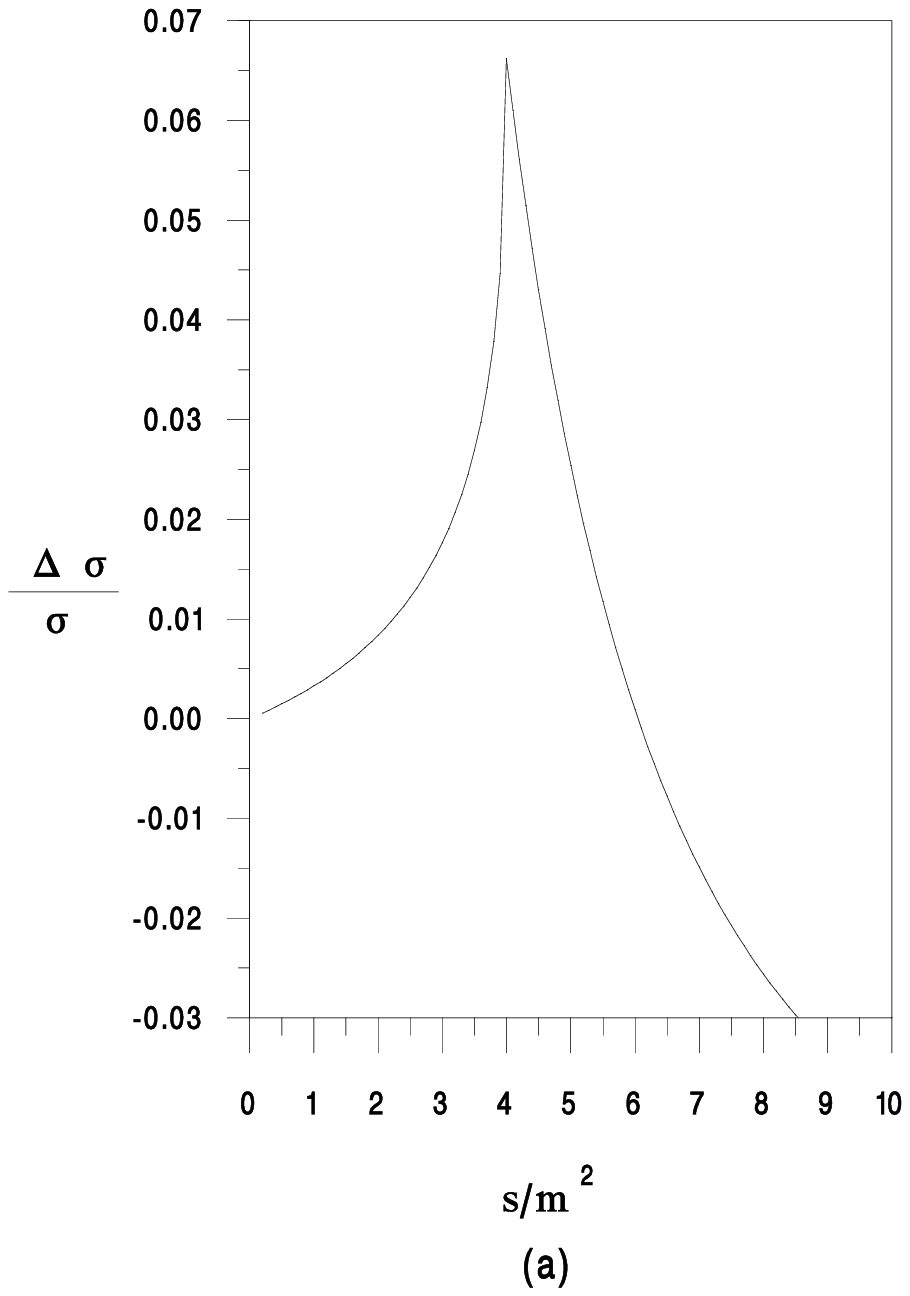,width=5cm}
\epsfig{figure=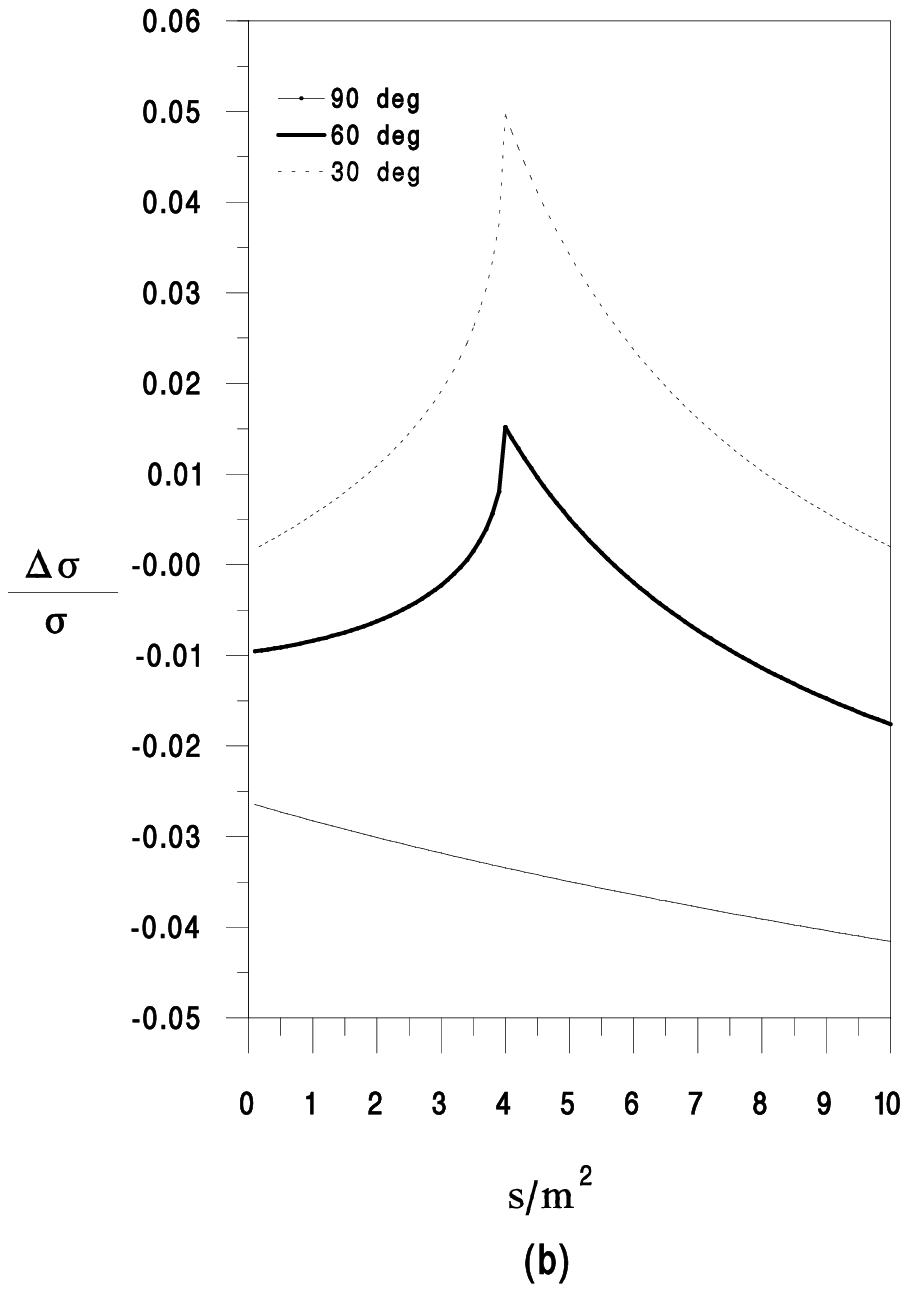,width=5cm}}\\
\mbox{\epsfig{figure=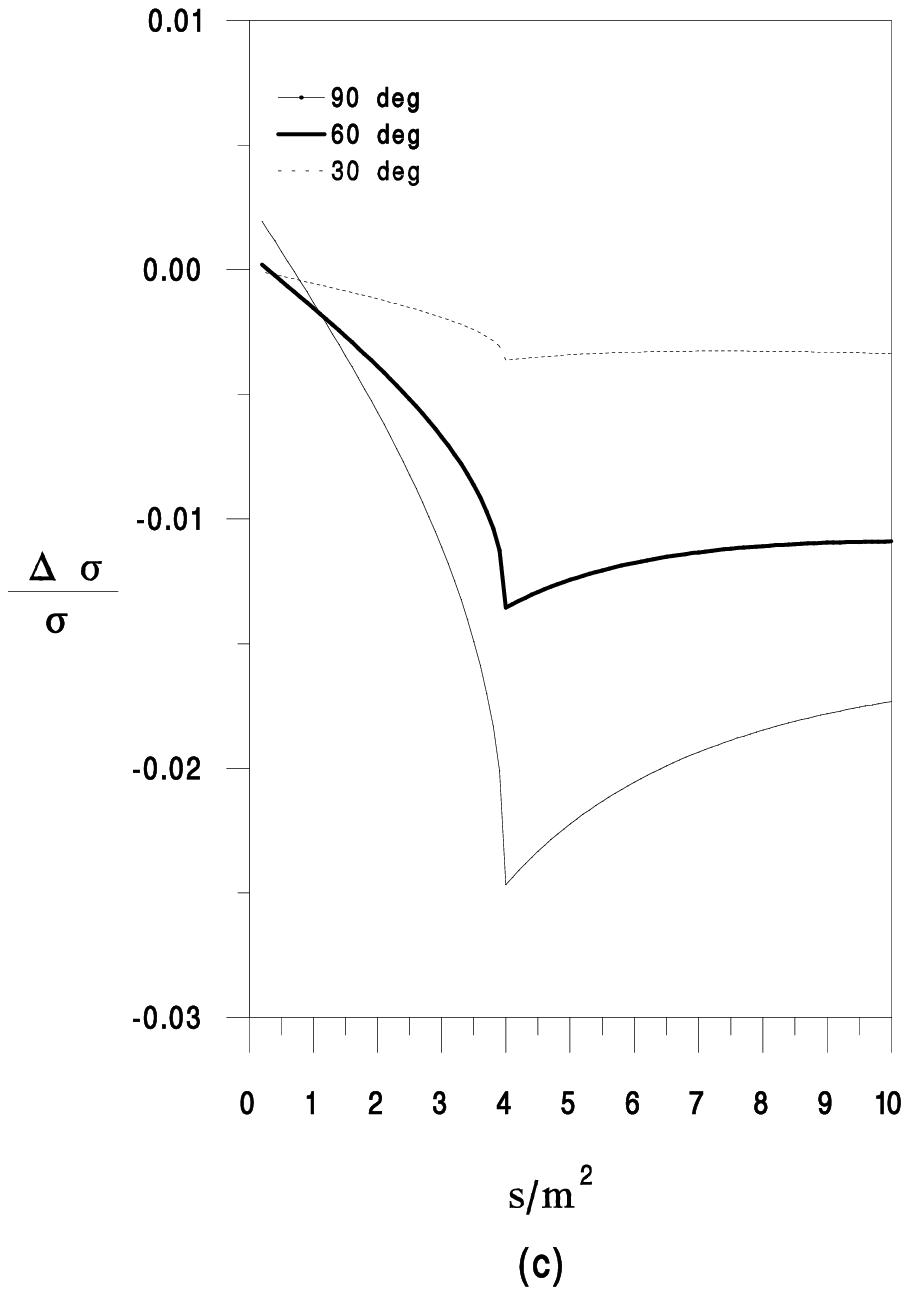,width=5cm}
\epsfig{figure=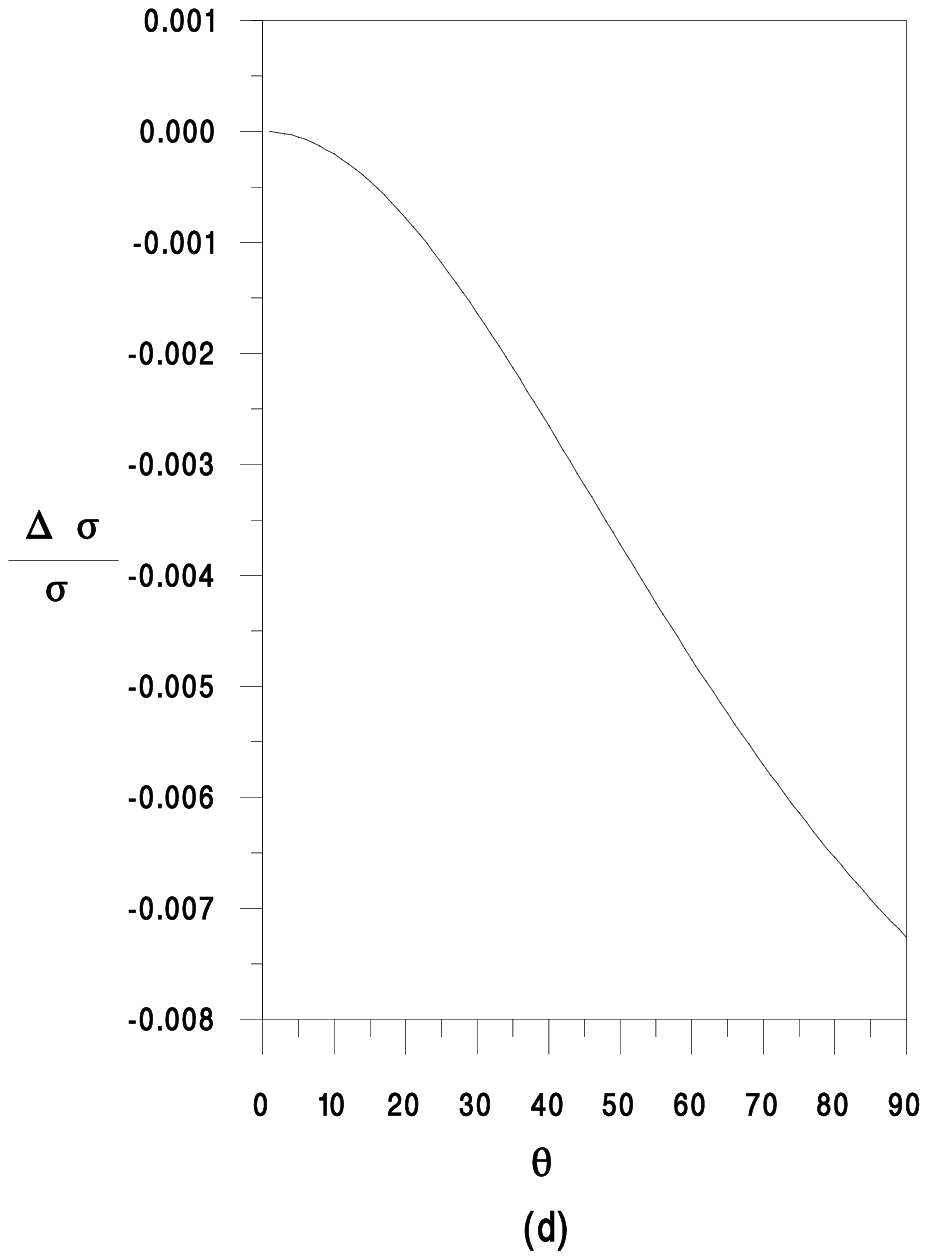,width=5cm}}
\caption[]{ 
One-loop virtual-sparticle corrections \cite{ER} in the threshold region of the
subprocess centre-of-mass energy squared $s$ to the processes (a) $q_j\bar
q_j\rightarrow q_k\bar q_k$, (b) $q\bar q\rightarrow gg$ for three different
subprocess centre-of-mass scattering angles, (c) $qg\rightarrow qg$ also for three
different values of scattering angle, and (d) $q_jq_k\rightarrow q_jq_k$, against
the centre-of-mass subprocess scattering angle, $\theta$, for $s = 10 m^2$. All
corrections are evaluated using $\alpha_s$ = 0.11.  
}
\end{figure}

\par Meanwhile, uncertainties in the parton distributions have been studied
carefully by two groups: one concluded that modifications in the distributions
of quarks inside protons could not explain away the CDF anomaly~\cite{Durham},
whereas the other group found that it could be removed by possible modifications
in the gluon  distribution~\cite{CTEQ}. This group also found that the CDF and D0
large-$E_T$ date could be reconciled once the different angular acceptances of
the two experiments were taken into account, as seen in Fig.~7. It seems,
therefore, that this anomaly has also found an interpretation within the
Standard Model.

\begin{figure}
\hglue1cm
\epsfig{figure=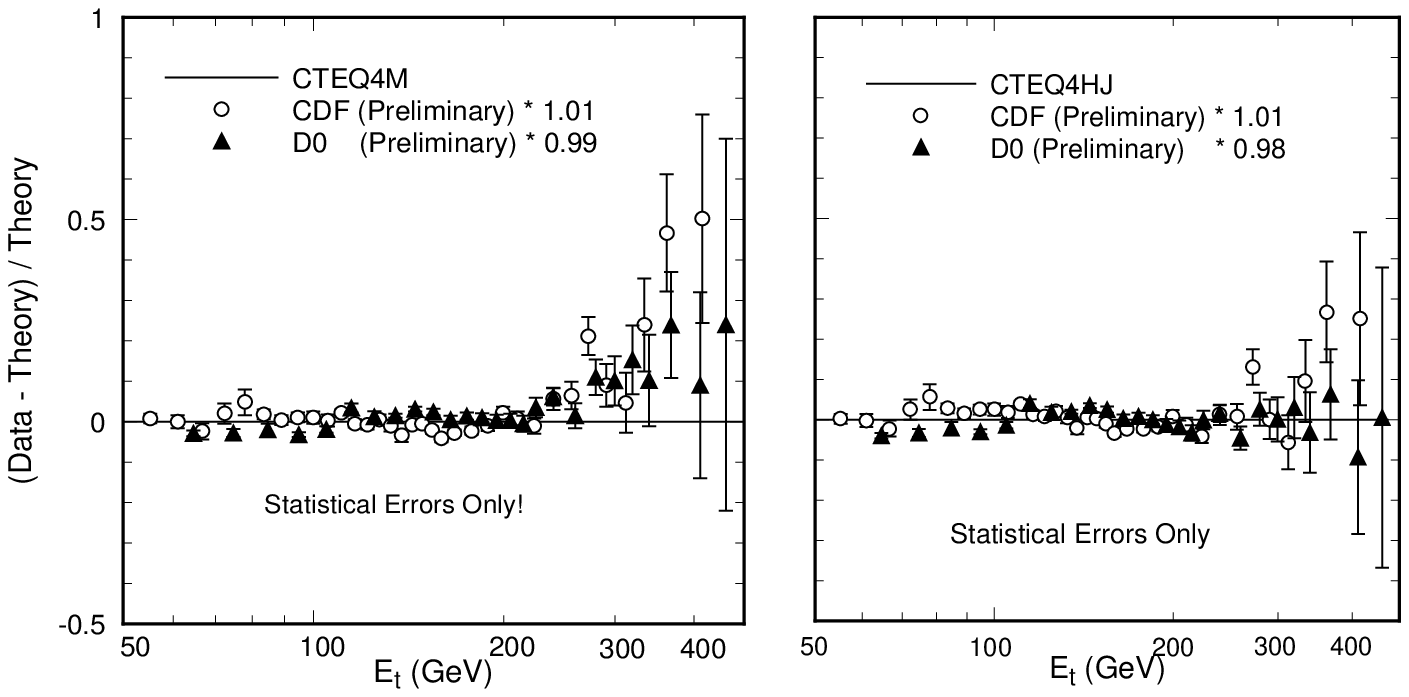,width=10cm}
\caption[]{
Large-$E_T$ jet data from CDF and D0 are compared with QCD predictions
based on recent parton distributions~\cite{CTEQ}.}
\end{figure}

\section{Grand Unified Theories}

We have already seen how the measurements of the different gauge couplings from
LEP and elsewhere are qualitatively consistent with the predictions of GUTs,
particularly those with supersymmetry~\cite{susygut}. This is fine, but what one
would really like to see is some evidence for one of the novel phenomena
predicted by GUTs, such as proton decay or neutrino masses.

\par The start-up of the Superkamiokande detector will enable existing lower
limits on the proton lifetime to be improved by an order of magnitude or
more~\cite{superk}. There is no guarantee that this will be sufficient, but at
least different GUT models such as minimal supersymmetric
$SU(5)$, missing-partner $SU(5)$ models and flipped $SU(5)$ make characteristic
predictions for the favoured decay modes~\cite{pdk}.  Thus, if proton decay is
seen, models will be distinguishable.

\par The available upper limits on neutrino masses are shown in Table~1: these
are so far below the masses of the corresponding charged leptons that one might
wonder whether they might vanish entirely. However, theoretically there is no
good reason, such as an exact gauge symmetry, why neutrino masses should be
strictly zero. Indeed, the consensus among GUT models is that neutrinos should
have masses, though the precise form of their mass matrix is very
model-dependent. One favoured possibility is the see-saw mass
matrix~\cite{seesaw}:
\beq
(\nu_L,\bar\nu_R)~~\left(\matrix{\sim 0 & \sim m_q\cr \sim m_q & \sim
M_{GUT}}\right)~~\left(\matrix{\nu_L\cr \bar\nu_R}\right)
\label{seesaw}
\eeq
 where $\nu_{L,R}$ denote the usual electroweak doublet left- and
singlet right-handed neutrinos, respectively. Diagonalizing (\ref{seesaw}) leads
to very light (very nearly) left-handed neutrinos with masses:
\begin{equation} m_{{\nu}_i} \, \simeq \, {m^2_{q_i} \over M_{GUT}} \, \gg \,
m_{q, \ell}
\label{numass}
\end{equation} By analogy with the Cabibbo-Kobayashi-Maskawa mixing of the
$W^{\pm}$ couplings to the quarks, one also expects mixing between the different
neutrino species, which leads in general to neutrino oscillations.

\par Neutrino oscillations rank among the possible interpretations of three
anomalous phenomena in neutrino experiments. Foremost among these is the {\bf
Solar Neutrino Deficit} now seen in five experiments~\cite{solarnu}. These
results cannot be explained away by {\it ad hoc} modifications of the Standard
Solar Model, e.g., by simply postulating a reduction in the central temperature 
of the Sun, without running into trouble with (for example) helioseismological
observations~\cite{helio}. On the other hand, there are three candidate
neutrino-oscillation interpretations: in  terms of matter-enhanced
Mikheyev-Smirnov-Wolfenstein~\cite{MSW} oscillations with a small mixing angle
$\theta$:
\begin{equation}
\Delta m^2_{\nu} \simeq 10^{-5} \hbox{eV}^2, \, 
\hbox{sin}^2 2 \theta \simeq 10^{-2},
\label{mswsmall}
\end{equation} or with a large mixing angle:
\begin{equation}
\Delta m^2_{\nu} \simeq 10^{-5} \hbox{eV}^2, \,
\hbox{sin}^2 2 \theta \simeq 1,
\label{mswbig}
\end{equation} or non-enhanced oscillations {\it in vacuo}:
\begin{equation}
\Delta m^2_{\nu} \simeq 10^{-10} \hbox{eV}^2, \,
\hbox{sin}^2 2 \theta \simeq 1.
\label{vacuum}
\end{equation} These different interpretations may be distinguished by new
experiments now starting to take data and in preparation, notably the
Superkamiokande, SNO and Borexino experiments~\cite{newsolarnu}.

\par Of the three interpretations presented above, (\ref{mswsmall}) seems the
most plausible from the point of view of the see-saw mechanism (\ref{seesaw}),
with
\begin{equation} m_{\nu_e} << m_{\nu_\mu} \simeq 3 \times 10^{-3} \hbox{eV}
\label{elessmu}
\end{equation} Furthermore, if one scales (\ref{elessmu}) up to the $\nu_{\tau}$
mass as in (\ref{numass}), this interpretation carries with it the intriguing
speculation that
\begin{equation} m_{\nu_{\tau}} \simeq ({m_t^2 \over m_c^2}) \times m_{\nu_{\mu}}
\simeq (1 - 10) \hbox{eV}
\label{nutaumass}
\end{equation} in which case the $\nu_{\tau}$ could provide the Hot Dark Matter
of interest to cosmologists~\cite{HDM}.  If $m_{\nu_{\tau}}$ is in the range
(\ref{nutaumass}), $\nu_{\mu} - \nu_{\tau}$ oscillations might be within reach
of accelerator neutrino experiments. Two suitable experiments are now taking
data at CERN~\cite{nuosc}, and another is planned at Fermilab~\cite{COSMOS}.

\par An {\bf Atmospheric Neutrino Deficit} in the ratio of $\nu_{\mu}$- to
$\nu_e$-induced events has been reported by some underground detectors, notably
the Kamiokande and IMB experiments~\cite{atmosnu}. However, this has not been
confirmed by other experiments using different techniques, and requires further
confirmation.

\par Also in need of confirmation is the suggestion of $\nu_{\mu}
\rightarrow \nu_e$ oscillations by the {\bf LSND Experiment}~\cite{LSND}, which
may be forthcoming from the KARMEN experiment~\cite{KARMEN} at the
Rutherford-Appleton Laboratory.

\section{LHC}

Some of the best prospects for addressing key current issues in the
phenomenology of particle physics will be provided by the LHC accelerator, which
has been approved for construction  at CERN. This is planned to collide protons
at a centre-of-mass energy of $14$ TeV with a luminosity of $10^{34}$ 
cm$^{-2}$s$^{-1}$, and lead ions at $1.2$ PeVin the centre of mass with a
luminosity of $10^{27}$ cm$^{-2}$s$^{-1}$. The $pp$ option is primarily intended
to address the Issue of Mass, by producing and detecting the Higgs boson and
supersymmetric particles, if they exist, whereas the lead-lead option is aimed at
the production and detection of the quark-gluon plasma.

\par As seen in Fig.~8, it is believed that  experiments at the LHC can detect
the Higgs boson of the Standard Model if it weighs above $90$ GeV, thus
dovetailing nicely with LEP $2$. The $H \rightarrow \gamma \gamma$ decay mode
should be detectable if $90 \hbox{GeV} < M_H < 150$ GeV, and Higgs decay into
two real or virtual $Z$ or $W$ bosons if
$120 \hbox{GeV} < M_H < 1$ TeV. As seen in Fig.~9, the Higgs bosons of the
minimal supersymmetric extension of the Standard Model should be detectable over
essentially all of the model's parameter space~\cite{Richter}.

\begin{figure}
\hglue2cm
\epsfig{figure=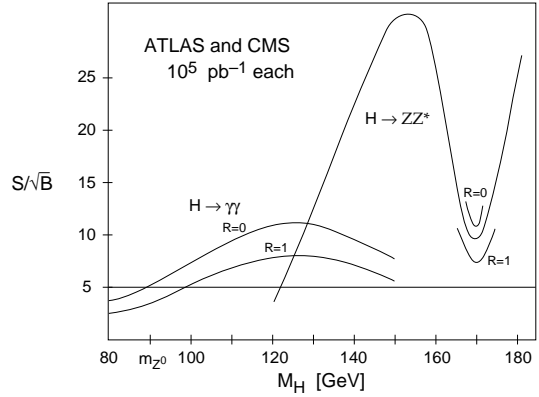,width=7cm}
\caption[]{
The expected significance for a Standard Model Higgs boson in the
ATLAS and CMS experiments at the LHC \cite{ATLASCMS}. The higher-mass range
is also accessible up to about 1~TeV.}
\end{figure}

\begin{figure}
\hglue2cm
\epsfig{figure=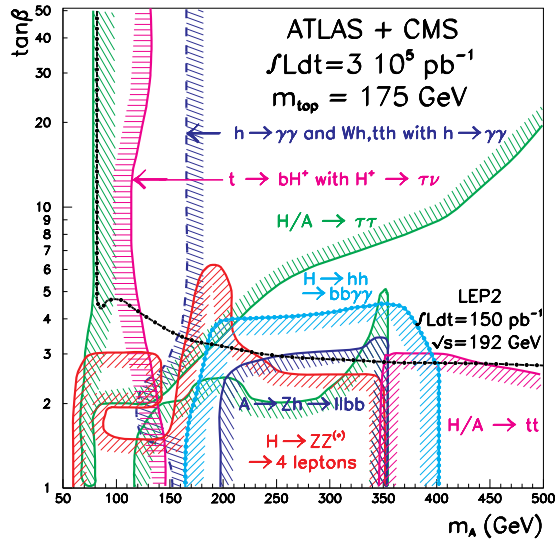,width=8cm}
\caption[]{ 
Capability of LHC experiments to explore the MSSM
Higgs sector~\cite{Richter}. The regions with shaded edges can be explored with
the channels indicated. Also shown is the region accessible to LEP2. Between LHC
and LEP2, essentially the entire plane is covered.
}
\end{figure}

\par Intense efforts are now underway to evaluate precisely the ability of the
LHC experiments to detect the strongly-interacting squarks and gluinos. Some
recent results~\cite{Paige} are displayed in Fig.~10: the missing-energy
signature for sparticle decay stands out above the total standard Model
background and  possible detector effects. Preliminary studies indicate that
experiments at the LHC can detect squarks and gluinos weighing up to $1 - 2$
TeV~\cite{ATLASCMS}, covering the entire range (\ref{susynear}) suggested by the
naturalness of the mass  hierarchy. It also seems that the LHC may be able to
provide us with some precision measurements of sparticle masses~\cite{Paige}

\begin{figure}
\hglue3cm
\epsfig{figure=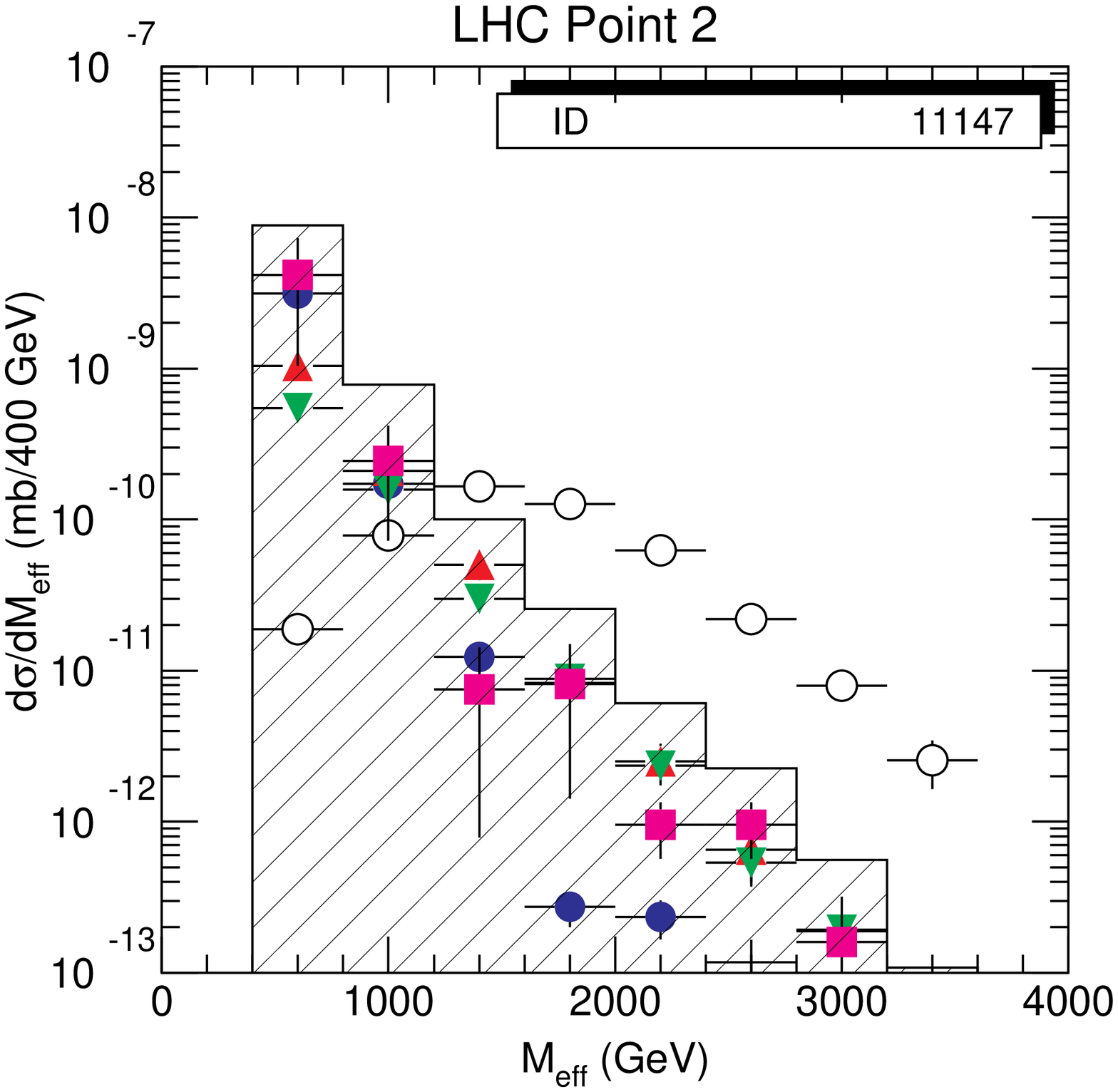,width=5cm}
\caption[]{ 
Comparison of calculated missing-energy signature due to ${\tilde
q}, {\tilde g}$ production at the LHC with Standard Model and
detector backgrounds~\cite{Paige}, in a model with dynamical EWSB and $m_0$ = 
400 GeV, $m_{1/2}$ = 400 GeV, $\tan\beta$ = 10 and $\mu > 0$.}
\end{figure}

\section{Beyond the Standard Model?}

\par We have seen in previous sections that there are good theoretical reasons
to expect new physics beyond the Standard Model, and that there are some
tentative experimental indications that may point towards the way to go. I
attach particular significance to the indications from precision electroweak
measurements at LEP and elsewhere that the Higgs boson may be relatively light
as suggested by supersymmetric models, the agreement between the measurements of
gauge couplings and the predictions of simple supersymmetric GUTs, and the
apparent solar neutrino deficit. In my view, these are pointers towards
supersymmetry and GUTs. 

\par We are fortunate that experiments now starting to take data  should be able
to explore broad domains of parameter space. Specifically, I have in mind the
CHORUS and NOMAD neutrino oscillation experiments at CERN~\cite{nuosc}, LEP $2$,
and the Superkamiokande underground detector~\cite{superk}. Beyond these, we
have  in prospect many further neutrino oscillation experiments and the LHC to
spearhead our search for new physics. Ever since the establishment of the
Standard Model, we have been waiting in vain for physics beyond the Standard
Model to appear. Will our luck soon improve?

\end{document}